\documentclass{article}

\usepackage{graphicx}
\graphicspath{{./eps/}}
\DeclareGraphicsExtensions{.eps}

\usepackage{color}
\usepackage{amsmath,amsfonts}
\usepackage{mathdots}
\usepackage{verbatim}
\usepackage{subfigure}
\usepackage{algorithm,algorithmic}

\begin{document}


\title{Using Linear Dynamical Topic Model for Inferring Temporal Social Correlation in Latent Space}
\author{Freddy Chong Tat Chua, Richard J. Oentaryo, and Ee-Peng Lim\\
Living Analytics Research Centre\\
School of Information Systems\\
Singapore Management University\\
80 Stamford Road, Singapore 178902}







\date{}

\maketitle

\begin{abstract}
	The abundance of online user data has led to a surge of interests in understanding the dynamics of social relationships using computational methods. Utilizing users' items adoption data, we develop a new method to compute the Granger-causal (GC) relationships among users. In order to handle the high dimensional and sparse nature of the adoption data, we propose to model the relationships among users in \emph{latent space} instead of the original data space. We devise a \emph{Linear Dynamical Topic Model} (LDTM) that can capture the dynamics of the users' items adoption behaviors in latent (topic) space. Using the time series of temporal topic distributions learned by LDTM, we conduct \emph{Granger causality} tests to measure the social correlation relationships between pairs of users. We call the combination of our LDTM and Granger causality tests as \emph{Temporal Social Correlation}. By conducting extensive experiments on bibliographic data, where authors are analogous to users, we show that the ordering of authors' name on their publications plays a statistically significant role in the interaction of research topics among the authors. We also present a case study to illustrate the correlational relationships between pairs of authors.
\end{abstract}

\section{Introduction}
\label{chap:gc:sec:motivation}

Rapid advances in social media and internet technologies have led to the generation of massive user data in digital forms. This gives rise to an important question: How do users relate to and socially influence one another? \emph{Social influence} is the mechanism of a user modifying her behavior or attributes so as to be more similar to her other socially connected users. For many decades, social scientists recognize the importance of social influence contributing to homophily in social networks, and have embarked on research that determine and measure the effect of social influence on homophily \cite{Kandel1978,Friedkin1998}. Measuring social influence has many practical applications; for instance, it provides an effective means to target influential individuals for product marketing, or to identify pivotal people in an organization for optimizing corporate management as well as driving innovations.

In this paper, we define social influence from a user $i$ to another user $j$ as ``the actions of $i$ causes $j$ to perform a set of actions in the future''. Social influence has been previously studied by various researchers \cite{Cui2011a,Liu2010a,Tang2009,Cheng2012}. However, their approaches do not take into account the temporal aspects of social influence. Instead of analyzing users' past and future actions, they consider user actions independent of their timestamps.


Many existing works also fail to account for the causality aspect of social influence. Knowing how user $i$'s past actions can predict user $j$'s future actions better than $j$'s past actions is only a necessary condition and not sufficient for finding social influence. Since the definition of social influence reflects the widely discussed notion of \emph{causality} \cite{Granger1980,Pearl2000}, the sufficient condition for finding social influence requires us to exclude other external factors that could affect the actions of $j$. That is, we need to eliminate the confounding variables that give doubt to the predictive power of $i$'s and $j$'s past on $j$'s future \cite{Granger2001}.

It is generally difficult, however, to satisfy this sufficient condition, due to the absence of complete user data capturing all external factors that influence the users' actions. There is also a need to conduct randomized controlled experiments \cite{greenland90,Rubi:1978}, which is very challenging in practice. Given these difficulties, we relax our assumptions and use a \emph{simplified notion} of social influence, known as \emph{Temporal Social Correlation} (TSC), whereby we ignore the presence of confounding variables, and assume that users who are socially correlated tend to make similar choices over time.

\subsection{Problem Formulation}

We apply the aforementioned notion of social influence for the analysis of users' items adoption behavior. We use the term ``users adopting items'' to refer to any action of a user on items reflecting her preferences. The concept of users adopting items can be applied in various contexts, e.g., users watching movies, users joining online communities \cite{Chua2011,Chua2012a}, or users producing words \cite{Chua2012}. 

In this research, we model social influence-driven changes in users' adoption behavior as a form of information transfer between users. We first obtain the time series representation of users' behavior using our proposed \emph{Linear Dynamical Topic Model} (LDTM), then we quantify the information transfer using \emph{Granger causality} (GC) tests \cite{Granger2001}, resulting in the derived \emph{Temporal Social Correlation} (TSC) values between two users. We say that ``$j$ follows $i$'' or ``$i$ transfers information to $j$'' when Temporal Social Correlation (TSC) exists from $i$ to $j$ at the time point of their interaction $\tau$; we denote this as $TSC(i \rightarrow j, \tau)$. With respect to our simplified notion of social influence, we shall hereafter use the term ``follow'', ``information transfer'' or ``Granger cause'' in place of ``social influence'', since causality cannot be proven adequately without randomized experiments. It is also worth noting that $TSC(i \rightarrow j, \tau)$ and $TSC(j \rightarrow i, \tau)$ are not necessarily the same.

\begin{figure}[htb]
	\centering
	\includegraphics[width=0.8\textwidth]{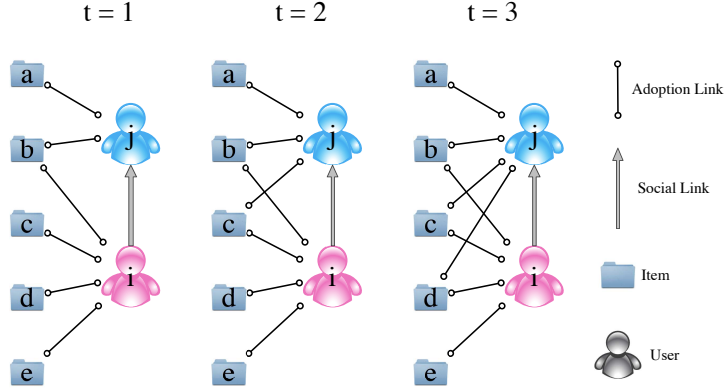}
	\caption{Example of Temporal Social Correlation from $i$ to $j$ in temporal adoption data}
	\label{chap:gc:fig:influence_example}
\end{figure}

As an illustration for Temporal Social Correlation in temporal item adoption data, consider the pedagogical example in Figure \ref{chap:gc:fig:influence_example}. The figure shows two users $i$ and $j$ adopting different subsets of five items over three time steps. When temporal information is missing, we could only observe the adoption states at the last time step (i.e., $t=3$), but based on the most recent states alone we cannot tell whether $i$ follows $j$ or $j$ follows $i$. Only by observing the adoption states of $t=1$ and $t=2$, we can infer that $j$ progressively follows $i$ in adopting item $c$ at $t=2$ and item $d$ at $t=3$. The converse is unlikely because $i$'s adoption states remain the same over time. In other words, $i$'s adoption states at $t=1$ is sufficient to predict her states for $t > 1$.

We can further generalize the example in Figure \ref{chap:gc:fig:influence_example}, and arrive at the following problem formulation: \emph{Given a set of users $U$ and a set of items $V$ that $U$ adopt from time step 1 to $T$, determine the $TSC(i \rightarrow j, \tau)$ and $TSC(j \rightarrow i, \tau)$ for all pair of users $i, j \in U$ when $i$ and $j$ interacts at a specific time point $\tau \in \{1,\ldots,T\}$. When $TSC(i \rightarrow j, \tau) > TSC(j \rightarrow i, \tau)$, we can say that $i$ influences $j$, or $j$ follows $i$.} 

\subsection{Measuring Causality in Adoption Data}

To quantify $TSC(i \rightarrow j, \tau)$ in item adoption data, one can take a straightforward approach, directly derived from the data. First, the raw frequencies of the adopted items for users $i$ and $j$ at time step $t$ can be represented as \emph{adoption vectors} $v_{i,t} \in \mathbb{R}^{M}$ and $v_{j,t} \in \mathbb{R}^{M}$ respectively, where $M$ is the total number of items. Vectors for each user $i$ over $T$ time steps form a time series $\{ v_{i,1}, \ldots, v_{i,T} \}$. (An additional normalization step, e.g., Term Frequency and Inverse Document Frequency (TF-IDF), may be performed a priori on the raw frequencies to balance the importance of popular and unique items.) Subsequently, one can compare the time series $\{ v_{i, 1}, \ldots, v_{i, T} \}$ and $\{ v_{j, 1}, \ldots, v_{j, T} \}$ and measure social influence by computing $TSC(i \rightarrow j, \tau)$ and $TSC(j \rightarrow i, \tau)$.

Despite the simplicity, this direct approach gives rise to several issues:
\begin{enumerate}
	\item The adoption vectors $v_{i,t}$ are usually \emph{high dimensional} in practice, i.e., the number of items $M$ is often large. As a result, comparing vectors $v_{i,t}$ and $v_{j,t}$ of two users $i$ and $j$ would be computationally demanding (even with a linear-time algorithm).
	

	\item A related issue is the \emph{sparse} nature of the adoption vectors $v_{i,t}$, since each user only adopts a small subset of items at a particular time $t$. Comparing two sparse vectors will hardly yield any indication of significant relationship between them, because we ignore the co-occurrences of different items adopted by the users.

	\item Since the adoption counts accumulate over time, the rate of change in $v_{i,t}$ relative to its previous time step $v_{i,t-1}$ will gradually decay and become marginally small. At this point, the time series representing the user $i$'s behavior becomes stagnant. As $TSC$ measures how users change their behavior due to other users, stagnant time series can hardly show any correlation effects among the interacting users.

	\item If the time series $\{ v_{i,1}, \ldots, v_{i,T} \}$ and $\{ v_{j,1}, \ldots, v_{j,T} \}$ of users $i$ and $j$ are observed for a long period (i.e., large $T$), their comparison may give a misleading conclusion that no influence exists, because the $TSC$ between the two users typically takes place within a specific time window.
\end{enumerate}

\subsection{Proposal and Contributions}

To address issues 1) and 2), there is a need for a temporal latent factor model that can induce from sparse and high dimensional data, a compressed latent representation of the adoption behaviors over time. The latent representation should also exhibit good semantic interpretability. To handle issue 3), one may learn the users' latent factors at each time step independently. However, such na\"{i}ve approach is biased towards the most recent information and subject to catastrophic ignorance of the past behaviors. Normally, a user does not change her behavior abruptly and there should be a smooth, decaying transition of the user's latent factors over time. 
In consideration of the necessity for smooth transition of users' latent factors, we develop a method to automatically estimate a set of \emph{decay parameters} for balancing between the importance of past and recent information. Finally, to address issue 4), we need to specify a time window for constraining the comparison period in which $TSC$ is measured.

To fulfill these requirements, we propose in this paper a novel \emph{Linear Dynamical Topic Model} (LDTM). The proposed model represents each user's adoption behavior as \emph{topic distribution} (i.e., latent factors) at different time steps, and the evolution of the topic distribution is captured using the concept of \emph{Linear Dynamical Systems} (LDS) \cite{Chua2013,Ghahramani1996,Roweis1999,Sun2012}. Based on the topic distributions learned by LDTM, we can then conduct Granger causality tests to determine the social influence between pairs of users.

Deviating from the traditional methods that operate on the original data space, the proposed LDTM provides a novel inductive approach facilitating discovery of social influence and causality in latent topic space. To the best of our knowledge, LDTM is also the first kind of dynamic topic model that comprehensively models the dynamics of the users' adoption behaviors by leveraging on the LDS concept. We summarize our key contributions as follows:
\begin{enumerate}
	\item Utilizing LDS to model the transition of topic distributions over time, we can automatically compute, for each user $n$, a dynamics matrix $A_{n, t}$ that contains the decay parameters of the user's adoption behavior at every time step. Such transition modeling via dynamics matrix $A_{n, t}$ has not been proposed in any topic models.


	\item For LDS estimation on the topic distribution parameters, we develop a forward inference algorithm based on the idea of \emph{Kalman Filter} (KF) \cite{Kalman1960}. The optimization of the dynamics matrix $A_{n, t}$ is done in such a way that maintains the notion of decaying adoption behavior, while ensuring that the temporal correlations of parameters in each topic distribution remains numerically stable.

	\item For inference of the decay parameters in $A_{n, t}$, we develop a new alternative method that aims at minimizing the \emph{Kullback-Leibler (KL) divergence} between the expected posterior distribution at time step $t$ and the expected prior distribution at time step $t$. This approach conforms nicely to the notion of smooth transition in the users' latent factors. In addition, it is computationally simpler and more efficient than traditional LDS methods, which first perform the \emph{Rauch-Tung-Striebel} (RTS) smoothing algorithm \cite{Rauch1965} for backward inference and then an additional optimization step to derive the dynamics matrix \cite{Ghahramani1996,Roweis1999}.

	\item Based on the temporal topic distributions derived by LDTM, we are able to identify information transfer between pairs of users by means of Granger causality tests. Through extensive experiments on bibliographic data, including DBLP and ACMDL datasets, we find evidences for Granger causality among the paper co-authors. Our statistical significance tests also reveal that the ordering of the co-authors' names plays a role in determining the information transfer among them.
\end{enumerate}


The remainder of this paper is organized as follows. In Section \ref{chap:related}, we first review several works related to our research. Section \ref{chap:gc:sec:gc_framework} discusses several desiderata in modeling temporal adoption data. The proposed LDTM is subsequently presented in Section \ref{chap:gc:sec:model}, followed by the procedure for the Granger causality test in Section \ref{chap:gc:sec:granger_causality}. Section \ref{chap:gc:sec:experiments} presents the experimental results and discussions using the bibliographic datasets. Finally, Section \ref{chap:gc:sec:summary} concludes this paper.

\section{Related Work}
\label{chap:related}

We first introduce the classical concepts on social influence in Section \ref{chap:related:sec:social_influence}, and present a review of the existing latent factor approaches for modeling temporal data in Section \ref{chap:related:sec:dynamic_factor}. We also cover in Section \ref{chap:related:sec:topic_influence} some related works that use some of latent factor modeling and influence concepts but in a significantly different way from our approach.


\subsection{Social Influence}
\label{chap:related:sec:social_influence}

To eliminate confounding variables for proving the existence of social influence, researchers use \emph{randomized experiments} that involve treatment and control groups. 
\cite{6035877} created a Facebook application to test whether broadcast or personalized messages have social influence on friends of a recruited user. 
\cite{bond201261} conducted experiments on Facebook users to study whether online political messages could influence the voting decisions of users. 
\cite{Muchnik09082013} studied how the votes of news articles affected the articles' discussions .

An alternative to randomized experiments is to perform \emph{quasi-experiments}. This approach is similar to the traditional randomized experiments, but lacks the element of random assignment to treatment or control. Instead, quasi-experimental designs typically allow us to control the assignment to the treatment condition, but using some criterion other than random assignment. 
\cite{Aral22122009} adapted \emph{matched sampling} technique in Yahoo! Messenger data to distinguish between influence and homophily in the adoption of a mobile service application (Yahoo! Go) . 
\cite{Anagnostopoulos:2008:ICS:1401890.1401897} proposed the \emph{shuffle test} to distinguish influence from homophily.

Research on social influence has revolved around the adoption of a \emph{single} item and satisfaction of the confounding condition. The research we pursue in this paper is different in several ways. First, we consider a set of items adopted by users instead of just a single item. Second, we propose LDTM to translate the high-dimensional set of items adopted by users into a low-dimensional temporal latent representation.

While existing works prove the existence of social influence in the adoption of item for users, the social influence is expressed as a discrete value that simply indicates presence or absence of influence. By contrast, we propose to use Granger causality measure to quantify the level of social influence between every pair of users, indicating how correlated their adoption behavior are over time.

\subsection{Temporal Latent Factor Models}
\label{chap:related:sec:dynamic_factor}

In general, there are two forms of temporal latent factor models. The first form seeks to obtain more accurate latent factors in the temporal domain by obtaining latent factors that globally approximate the observed data \cite{Ahmed2011,Blei:2006:DTM:1143844.1143859,Koren2009a,Koren2009,Sun2012,Wang2006,Xiong2010}. The other form known as \emph{online learning} focuses on the efficiency of handling real time streaming data by maximizing the likelihood of the latent factors to fit the observed data from the most recent time window only \cite{Agarwal2010,Canini2009,Cao2007,Gohr2009,Hoffman2010,Mairal2010,Saha2012,Wang2011a}. 


We note that the online learning models are extensions of latent factor models that are themselves not necessarily designed for dynamic data, but for efficient learning of new model parameters given new additional data. In this paper, we are concerned with modeling user behavioral data using some dynamic latent factor models, instead of the online learning of the dynamic models.

\cite{Blei:2006:DTM:1143844.1143859} proposed Dynamic Topic Model (DTM) for text documents. DTM was extended from the Latent Dirichlet Allocation (LDA) \cite{Blei2003a} to model the evolution of words within topics, i.e. words prominently used in a particular topic at a particular time step will be replaced by a different set of words at a later time. However, our requirement is slightly different. Instead of the evolution of topic-word distributions, we focus on the evolution of document-topic distributions.

The evolution of document-topic distributions has not been considered previously in LDA-based models (e.g. DTM), because LDA is mainly used for modeling text documents that remain static over time. Hence, DTM does not consider the evolution of users' behavior in the way we do. When we apply LDA for modeling users' behavior, the users replace the role of the documents, while the adopted items replace the words. In our work, we assume that topic-item distributions remain static over time while the human users' evolve their preferences over time. Since the generative process in DTM does not meet our temporal requirements, we are motivated to develop LDTM that extends static LDA by utilizing the concepts of Linear Dynamical System (LDS).

For modeling users' behavior, 
\cite{Ahmed2011} used an exponential decay function to model the decay of users' search intent on search engines. But they assume that the parameters of the decay function remain constant for all topics and all users. On the contrary, we assume that there is a decay parameter for each topic and that the decay parameters vary for each user. We aim to estimate the decay parameters automatically, which are representative of the users' temporal behavior.

To automatically determine the natural decay of each topic, 
\cite{Wang2006} proposed a non-Markovian approach that models the trend of topics evolution. The key idea is to associate additional Beta distribution with each topic in order to generate the time stamps of the words sampled from the topics. But this approach assumes that each topic is only relevant for each specific time period, and does not directly model the evolution of user behavior.


Latent factor models have also been widely used for collaborative filtering in recommendation tasks, and several researchers have proposed dynamic latent factor models for handling temporal data \cite{Koren2009a,Koren2009,Sun2012,Xiong2010}. However, these approaches have always been focused on predicting users' ratings on items, and so their models cannot be directly applied for modeling users' items adoptions. Nevertheless, owing to the similarity in the fundamental concept of dynamicity in latent factors, we give an overview of these works here.

\cite{Koren2009a,Koren2009} developed TimeSVD++ to address temporal dynamics through a specific parameterization with factors drifting from a central time. Koren assumed that users' item ratings remain static over time, since users do not rate the same items in different time periods. However, in item adoption scenario, users could adopt the same items at different time periods with different frequency.


\cite{Xiong2010} extended the factorization of users' item ratings from a static $\mathbb{R}^{M \times N}$ matrix to a $\mathbb{R}^{M \times N \times T}$ tensor, where $N$, $M$, and $T$ represent the number of users, items, and time steps respectively. Three sets of latent factors were derived from their tensor factorization method (rather than just two in matrix factorization case). The additional set of latent factor, known as the time latent factor, can be used to derive the temporal users' and items' latent factors from its multiplication. But such time latent factor assumes that the items' latent factor evolves over time in the same way as the users' latent factors. However, we require the items' latent factor to remain static, while allowing only the users' latent factors to change.

\cite{Sun2012} proposed Dynamic Matrix Factorization (DMF) which uses Linear Dynamical Systems (LDS) \cite{Ghahramani1996,Roweis1999,Shumway2006,Simon2006}. The centerpiece of this work is a dynamic state-space model that builds upon probabilistic matrix factorization in \cite{Salakhutdinov2008,Salakhutdinov2008b} and Kalman filter/smoothing \cite{Kalman1960,Rauch1965} in order to provide recommendations in the presence of process and measurement noises. Although the LDS component of DMF is able to model the evolution of users' behavior, the latent factors obtained by 
\cite{Sun2012} are not constrained to be non-negative. Hence, their approach is not able to provide intuitive interpretation on the preferences of users' adoption behavior.

To summarize, all these prior works fail to satisfy the following requirements for inferring temporal social dependencies between users: 1) They are not explicitly designed to model item adoption data. 2) They do not obtain non-negative latent factors for easy interpretation of the users' behavior. 3) They neither assume that users' behavior can decay over time nor show how the users' behavior can evolve over time.

We combine the LDA and LDS approaches to obtain LDTM. Our LDTM is able to model the users' items adoption data, obtain probabilistic (non-negative) latent factors for characterizing user behavior over time, and automatically infer the optimal decay parameters for each user at different time steps.

\subsection{Topic-based Influence Measures}
\label{chap:related:sec:topic_influence}

To model a set of items for diffusion or inferring influence, many authors \cite{Cui2011a,Dietz2007,Gerrish2010,Liu2010a,Liu2012,Tang2009,Tang:2013:CCI:2487575.2487691,VerSteeg2013,Wen2010,Weng:2010:TFT:1718487.1718520}. have also turned to the use of topic models. 
\cite{Cui2011a} proposed an influence matrix to suggest what items a user should share to maximize their individual influence in their own community. Their matrix measures influence between users and items while ours measure between users and users. Similar to \cite{Romero2011}, 
\cite{Weng:2010:TFT:1718487.1718520} extended PageRank \cite{Brin1998107} to include topic models in the computation of influence between users. 

\cite{Gerrish2010} extended dynamic LDA to identify the most influential documents in a scientific corpus. But the dynamic LDA assumes that the documents' latent factors evolve only with small perturbations while words' latent factors evolve over time. 
\cite{Gerrish2010}'s work differ from our approach, because we allow greater variability in users' (documents') latent factors and assume that items' (words') latent factors remain constant. 

A notable contender to our approach is the work proposed by 
\cite{VerSteeg2013}. They also use topic models to reduce the dimensionality of item adoptions, followed by analysis using an information theoretic measure of causality known as \emph{transfer entropy}. The algorithm to estimate transfer entropy is based on the nearest neighbor approach developed in Statistical Physics \cite{Kozachenko1987,Kraskov2004,Victor2002}. But there are significant drawbacks to this approach. First, it makes no assumption on the joint distributions of the variables, and thus requires many time steps for achieving accurate estimation. It also ignores the temporal correlations between users' topic distributions and the users' behavior evolution. 

Apart from \cite{Gerrish2010,VerSteeg2013}, all the prior works which uses latent factors does not use the time information when inferring influence. In this aspect, our work goes beyond the norms by considering temporal users' items adoption and proposing several temporal models. We distinguish our work by devising a Linear Dynamical System (LDS) approach to linearly correlate the users' topic distributions, and using Granger causality that likewise assumes linear relationship among variables. Due to this linear assumption in Granger causality measure, our method also requires less number of time steps to derive an accurate measure of social influence between (pairs of) users.

\section{Desiderata in Modeling Temporal Adoption Data}
\label{chap:gc:sec:gc_framework}

Measuring $TSC$ between two users $i$ and $j$ requires two crucial steps. First, an accurate measure of the users' adoption behavior represented as time series vectors in latent space is required for every time step. That is, we require latent factors $\theta_{i,t}, \theta_{j,t} \in \mathbb{R}^{K}$ for each user pair $(i,j)$ at time step $t$. $\theta_{i,t}$ and $\theta_{j,t}$ has $K$ dimensions where $K$ is much smaller than $M$ (the total number of items). Second, a temporal correlation measure is needed to compare between the trends of two time series. Knowing how two time series temporally correlate should help us make better predictions or reduce our uncertainty for their future adoption behavior. However, we need to address some issues in modeling temporal adoption data, as elaborated in Sections \ref{chap:gc:sec:model_adoption} and \ref{chap:gc:sec:necessity_temporal}.

\subsection{Latent Representation of Temporal Adoption Data}
\label{chap:gc:sec:model_adoption}

We propose a new way of representing user's adoption behavior in temporal latent space as opposed to the traditional method of using only the frequency of adoption in high dimensional space. There are some advantages of representing adoption behavior in temporal latent space as well as some difficulties, which we will elaborate further.

\begin{figure}[htb]
  \centering
  \subfigure[$t=1$]
  {
    \includegraphics[width=2.6in]{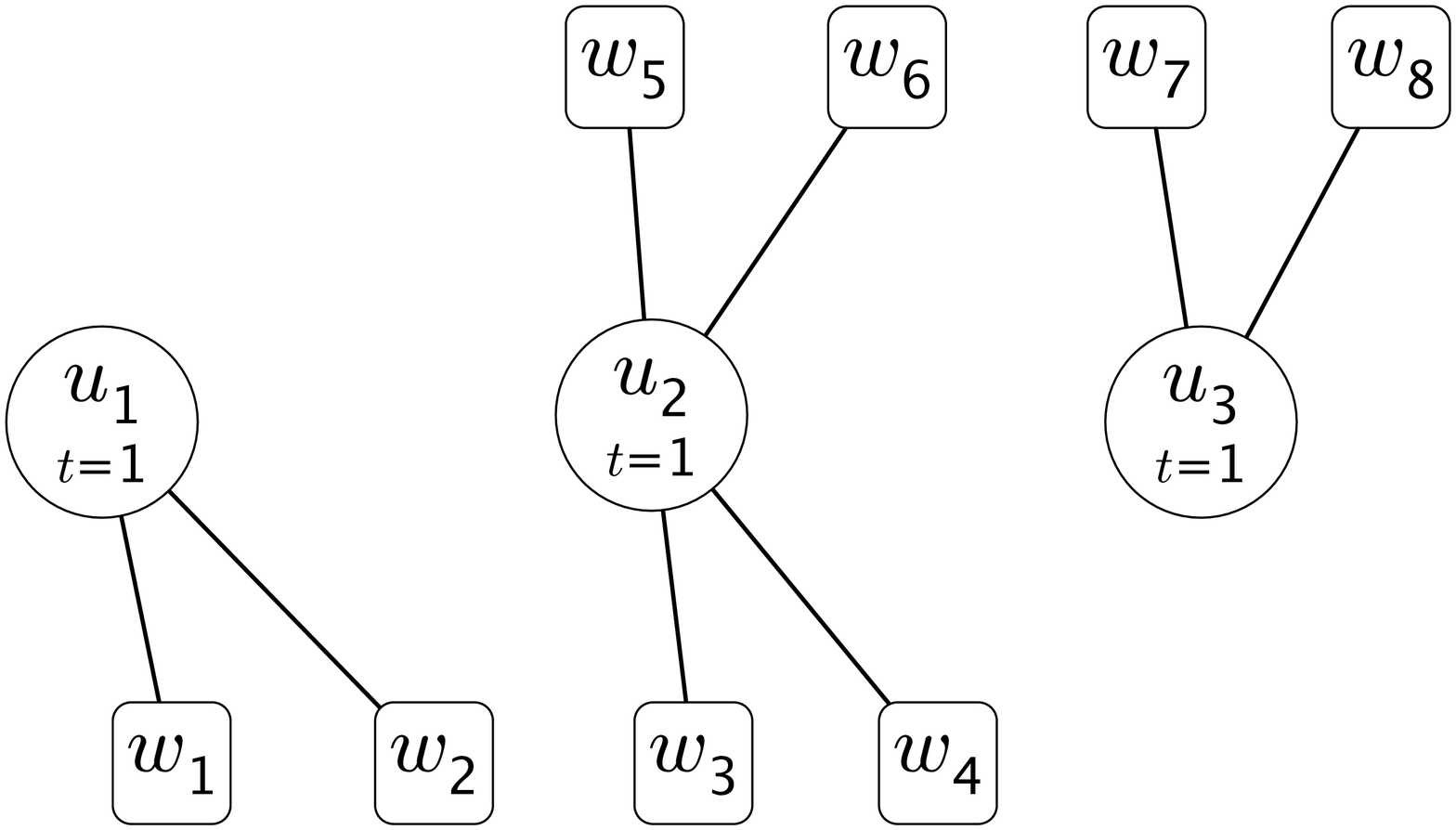}
    \label{chap:gc:fig:temporal_problem3}
  }
  \subfigure[$t=2$]
  {
    \includegraphics[width=2.6in]{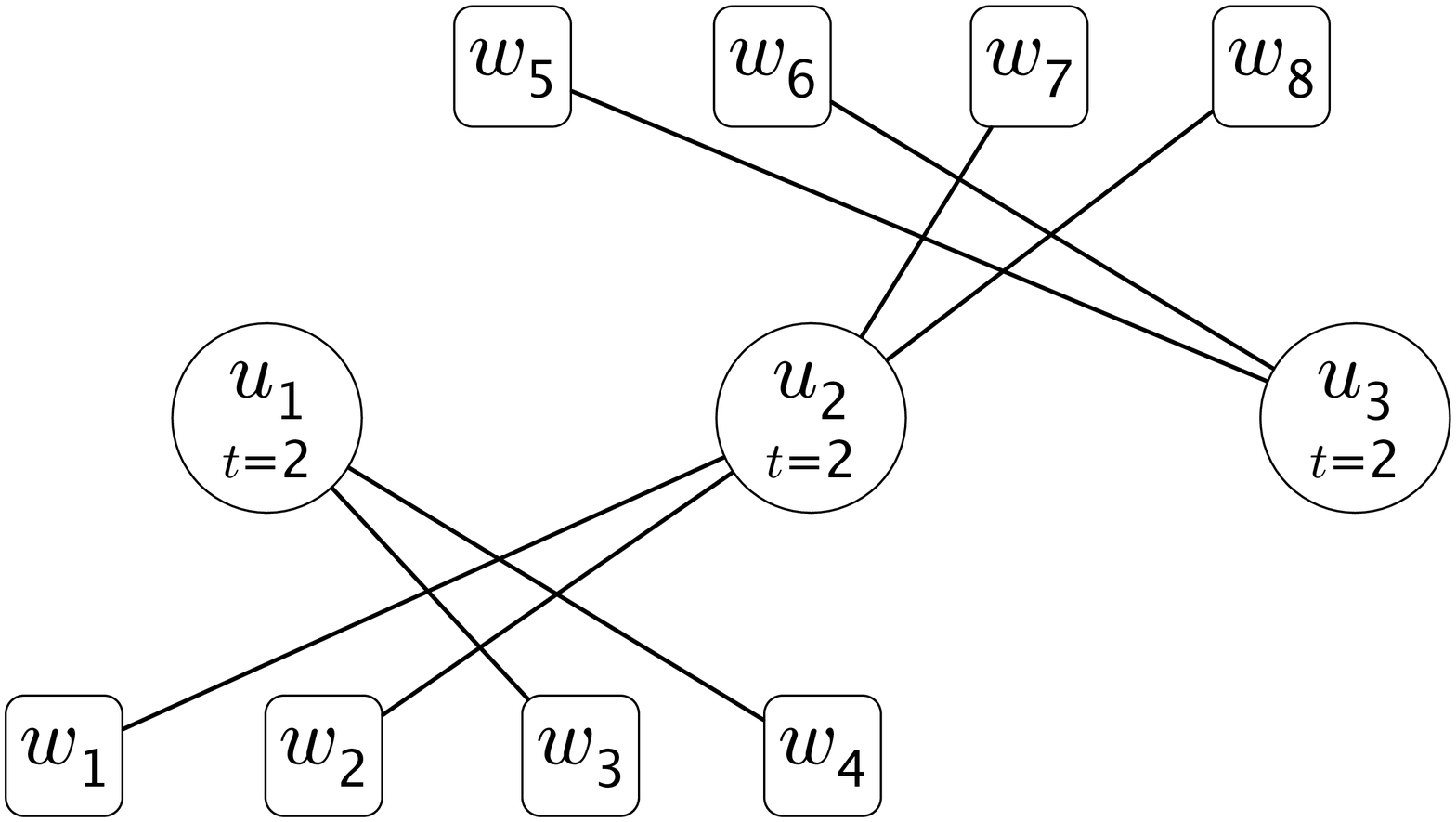}
    \label{chap:gc:fig:temporal_problem4}
  }
  \caption{Topic Modeling in Temporal User Item Adoptions}
  \label{chap:gc:fig:temporal_problem3n4}
\end{figure}


For illustration, consider the temporal item adoption problem in Figure \ref{chap:gc:fig:temporal_problem3n4}, involving three users $\{ u_1, u_2, u_3 \}$ and eight items $\{ w_1, \ldots, w_8 \}$ over two time steps. If we model the topic distributions at each time step independently of other time steps, we would obtain the scenarios in Figure \ref{chap:gc:fig:temporal_problem3} for time step 1, and Figure \ref{chap:gc:fig:temporal_problem4} for time step 2. One may see that the edges between users and items are sparse, which does not allow us to draw any meaningful intuitions about the relationship of items and does not show us any common item adoptions among the users.


\begin{figure}[htb]
  \centering
  \subfigure[Without time steps]
  {
    \includegraphics[width=2.6in]{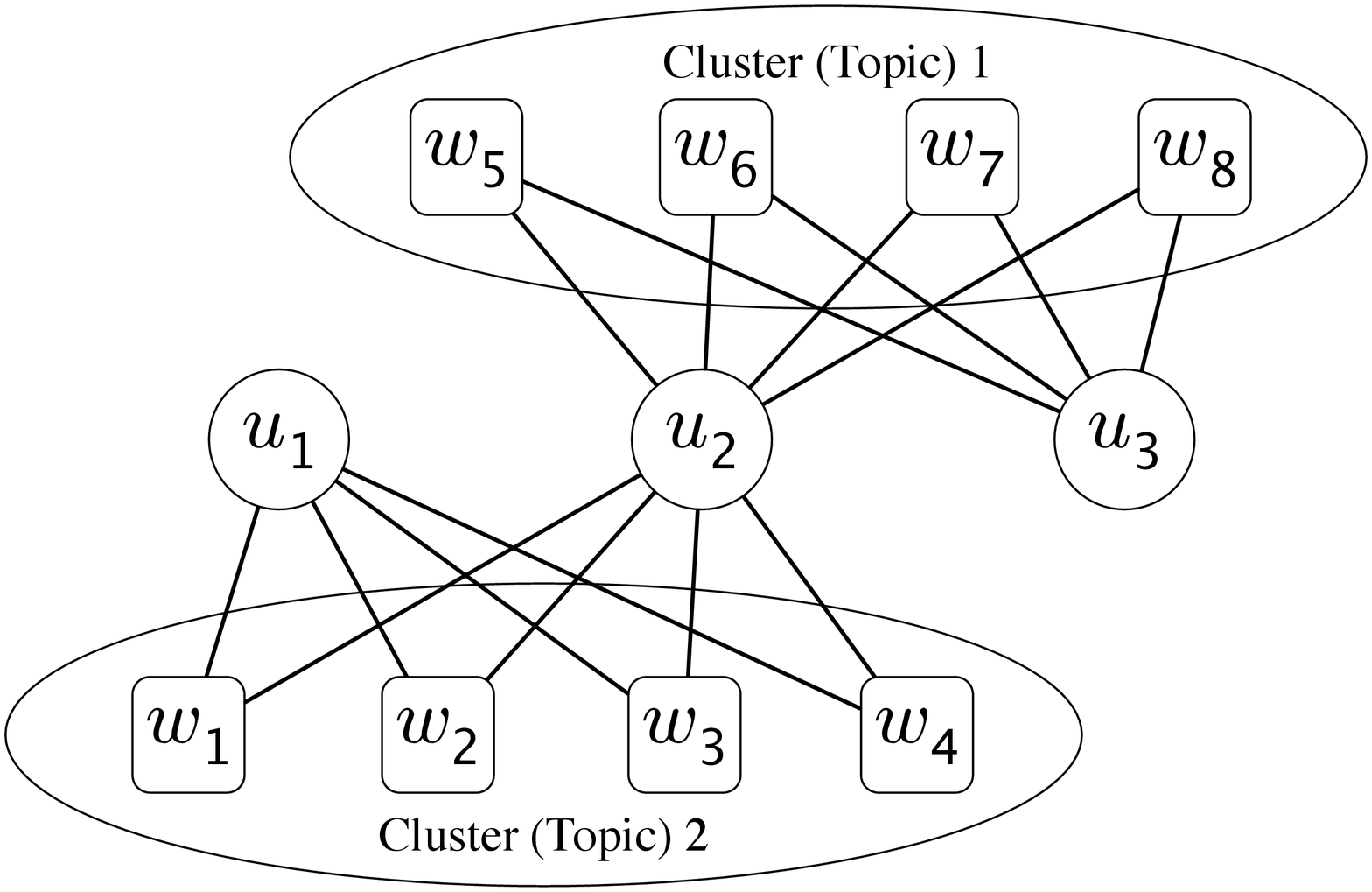}
    \label{chap:gc:fig:temporal_problem1}
  }
  \subfigure[With time steps]
  {
    \includegraphics[width=2.6in]{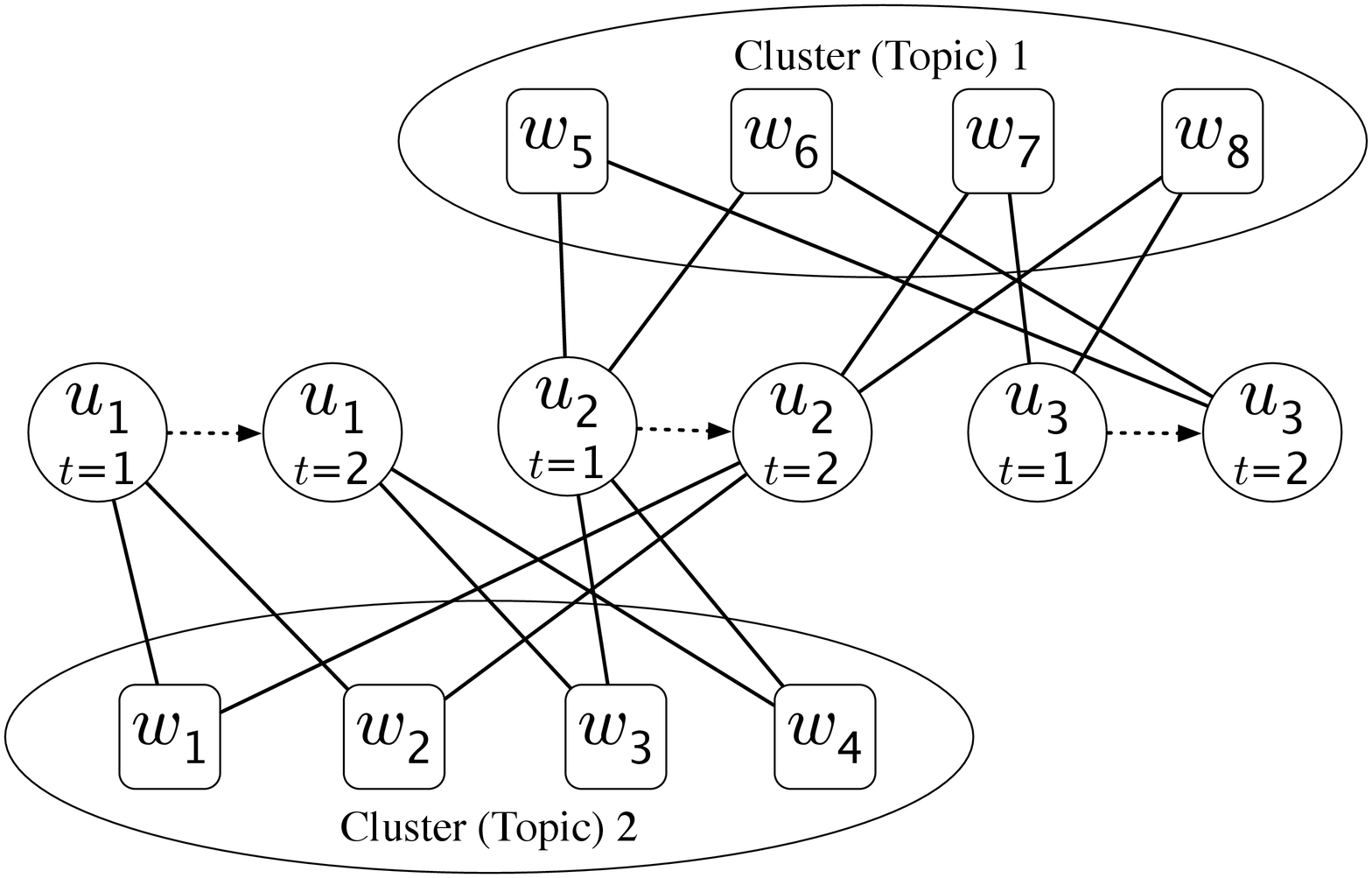}
    \label{chap:gc:fig:temporal_problem2}
  }
  \caption{Topic Modeling in Static User Item Adoptions}
  \label{chap:gc:fig:temporal_problem1n2}
\end{figure}


However, when we combine the temporal adoptions into a single time step, we obtain the scenarios as illustrated in Figure \ref{chap:gc:fig:temporal_problem1}. Figure \ref{chap:gc:fig:temporal_problem1} shows the result of performing topic modeling on data without temporal considerations. The items adopted by users $u_1, u_2$ and $u_3$ are clustered according to topics 1 and 2 based on the density of edges between users and items. We therefore require a method of modeling the temporal adoptions such that it allows us to preserve the edge densities across time steps and provides us with the topic distributions at different time steps. Such model could combine the temporal adoptions and construct dependencies between different time steps by having the scenario as shown in Figure \ref{chap:gc:fig:temporal_problem2}.


\subsection{The Need for Temporal Probabilistic Topic Model}
\label{chap:gc:sec:necessity_temporal}

There are many ways of modeling users' adoption behavior in latent spaces, and we wish to justify our choice of using probabilistic topic model. Besides probabilistic method, one may use Non-negative Matrix Factorizations (NMF) to obtain low-rank matrices that can substitute for the users' and items' latent factors \cite{Xu:2003:DCB:860435.860485,Liu2010}. Our previous work in temporal item adoptions has also explored the use of LDS with NMF \cite{Chua2013} for modeling evolving users' preferences. LDS with NMF can be stated as follows,
\begin{gather*}
  x_{n, t} = A_{n, t - 1} \cdot x_{n, t - 1} + \epsilon, \qquad \epsilon \sim \mathcal{N}(0, Q) \\
  w_{n, t, m} = C_{m} \cdot x_{n, t}
\end{gather*}
where $x_{n,t} \in \mathbb{R}^{K}$ is the vector representing user $n$'s adoption behavior at time step $t$, $A_{n,t-1} \in \mathbb{R}^{K \times K}$ is the dynamics matrix which evolves user's behavior from time $t-1$ to $t$, $w_{n,t,m} \in \mathbb{R}$ is the number of times user $n$ adopts item $m$ at time $t$, and $C_{m} \in \mathbb{R}^K$ represents item $m$'s latent factor.

In \cite{Chua2013}, we estimated the items latent factor matrix $C \in \mathbb{R}^{M \times K}$ for NMF by minimizing the sum-of-squared errors via stochastic gradient descent (SGD) with non-negativity constraints. The model was subsequently solved as an instance of Expectation Maximization (EM) algorithm \cite{Bilmes1997,Dempster1977}, where the E-step carries out Kalman Filtering and RTS Smoothing, and the M-step serves to optimize the dynamics matrix $A_{n, t}$.

In order to obtain \emph{interpretable topics}, it is imperative that the items latent factor matrix contains only \emph{non-negative values} \cite{Chua2013}. We often rank the importance of items according to the items' value in the respective latent factor, but this is not true if the latent factors contain negative values. A negative $c_{m,k}$ can also be important for contributing to the value $w_{n,t,m}$ if the corresponding $x_{n,t,k}$ is also negative. 

Due to the different amounts of item adoptions for each user at different time steps, a single static matrix $C$ that is defined in real space $\mathbb{R}^{M \times K}$ does not fit well for the adoption patterns of every user. $C$ was also only estimated once before running EM algorithm to estimate the rest of the parameters.


There is thus a strong requirement to have a \emph{non-negative} items' latent factor matrix that is \emph{normalized across different time steps} which is estimated by an algorithm that updates the items' latent factor \emph{iteratively} while learning the other parameters. Probabilistic approaches give us normalized parameters that sum to one and are non-negative (since probabilities cannot be less than zero). By alternating Gibbs Sampling with Kalman Filter and additional optimizing steps to derive the dynamics matrix, we derive an algorithm summarized in Algorithm \ref{chap:gc:alg:model} to estimate all the necessary parameters that achieves an overall better fit to the observed data.


\section{Linear Dynamical Topic Model}
\label{chap:gc:sec:model}

\begin{figure}[htb]
  \centering
  \includegraphics[width=0.50\textwidth]{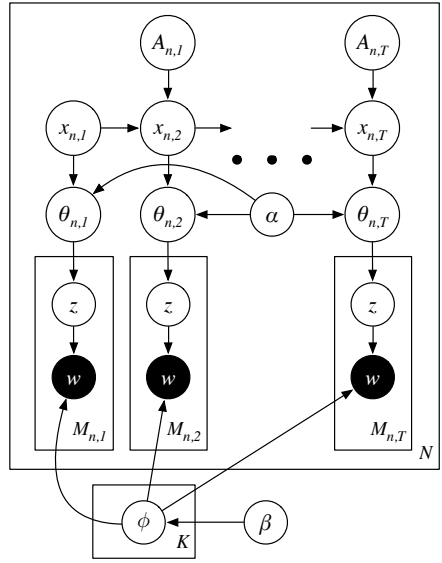}
  \caption{Probabilistic graphical representation of the proposed LDTM}
  \label{chap:gc:fig:plate}
\end{figure}

Figure \ref{chap:gc:fig:plate} shows the probabilistic graphical representation (a.k.a. Bayesian network) of LDTM using plate diagram. In essence, LDTM is a combination of Latent Dirichlet Allocation (LDA) and Linear Dynamical System (LDS). We obtain the users-topic distributions at each time step by inferring the latent topic variable conditioned on the words written in each time step and the topic-item distributions. 

\subsection{Modeling Assumptions}

We assume that the topic item distribution remains static over time, while the users' topic distribution evolves over time through a linear dynamical process conditioned on the previous time steps and the inferred latent variables in current time step. We further elaborate our assumptions of LDTM as follows:
\begin{enumerate}

  \item Given that there are $K$ topics and temporal adoption data, the topic distribution $\theta_{n,t}$ of user $n$ at time step $t$ is defined by the Dirichlet distribution with parameters $x_{n, t} \in \mathbb{R}^K$.
    \[ \theta_{n, t} \sim Dir( x_{n, t} ) \]

  \item To relate the current parameters $x_{n, t}$ with the previous time step parameters $x_{n, t - 1}$, we assume a linear distribution as defined by,
  \[ x_{n, t} = A_{n, t - 1} \cdot x_{n, t - 1} \]
  where $A_{n, t} \in \mathbb{R}^{K \times K}$ represents the dynamics matrix of user $n$ at $t$. This step distinguishes our model from all other topic models, i.e., we model the evolution of users' topic distribution using a dynamics matrix. We also derive a whole new set of inference equations for estimating the model parameters in Section \ref{sec:estimate_dynamics}.

  \item The topic $z_{n, t, m}$ of an item $m$ adopted by user $n$ at time $t$ is given by,
    \[ z_{n, t, m} \sim Mult( \theta_{n, t} ) \]
    Each topic item distribution is given by a simple symmetric Dirichlet distribution,
    \[ \phi_{k} \sim Dir( \beta ) \]
    Then each item $m$ adopted by user $n$ at time $t$ conditioned on topic variable $z_{n, t, m}$ is given by,
    \[ \left[ w_{n, t, m} | \left( z_{n, t, m} = k \right) \right] \sim Mult( \phi_{k} ) \]
\end{enumerate}

\subsection{Estimating Topic Distribution Parameters}

To calculate $TSC$, we require the topic distributions for each user $n$ at each time step $t$ conditioned on the information up to $t$ as denoted by $\theta_{n,t | t}$, also known as the posterior topic distribution. Since we have defined $\theta_{n, t}$ as a Dirichlet distribution with parameters $x_{n, t}$, knowing $x_{n, t | t}$ is sufficient for deriving $\theta_{n, t | t}$. $\theta_{n, t | t}$, the posterior topic distribution of user $n$ at time $t$ conditioned on information up to time step $t$ is given by,
\[ \theta_{n, t | t} \sim Dir( x_{n, t | t} ) \]
$x_{n, t | t}$, the posterior parameters of the Dirichlet distribution for user $n$ at time $t$ conditioned on information up to time step $t$ is given by a slight modification of the \emph{Kalman Filter} \cite{Kalman1960} algorithm,
\[ x_{n, t | t} = x_{n, t | t - 1} + \psi_{n, t} \]
where $\psi_{n, t} \in \mathbb{R}^K$ and $\psi_{n, t, k}$ denote the number of times user $n$ at time $t$ generated topic $k$. $x_{n, t | t - 1}$ is the prior parameters of the Dirichlet distribution for user $n$ at time $t$ conditioned on information up to time step $t-1$,
\[ x_{n, t | t - 1} = A_{n, t - 1} \cdot x_{n, t - 1 | t - 1} \]
where $A_{n, t-1} \in \mathbb{R}^{K \times K}$ is the dynamics matrix that evolves the parameters from $t-1$ to $t$. If $A_{n, t}$ for all time steps $t$ is assumed to be an identity matrix, the model reduces to the traditional LDA model for temporal data sets.

In previous works that use LDA on static data, there is a lack of emphasis on the importance of posterior and prior distributions. In temporal data, it is more important to distinguish between the two, as the posterior parameters of time step $t-1$ becomes the prior parameters at time step $t$ after factoring in the dynamics matrix $A_{n, t-1}$.

\subsection{Estimating the Decay Parameters for Dynamics Matrix}
\label{sec:estimate_dynamics}

Since the dynamics matrix $A_{n, t-1}$ gives us the prior distribution $\theta_{n,t|t-1}$, an ideal dynamics matrix should be able to predict the posterior distribution $\theta_{n,t|t}$ well. Therefore, to find the optimal dynamics matrix would require us to minimize the divergence between the expected prior and expected posterior distribution. A simple divergence to use would be the Kullback-Leibler (KL) Divergence. We minimize the KL Divergence between the expected posterior topic distribution $\theta_{n,t|t}$ and expected prior topic distribution $\theta_{n,t|t-1}$ of user $n$ at time $t$ as follows,
\[ \textit{minimize} \quad \sum_{t=2}^T D_{KL} \left[ E \left( \theta_{n,t|t} \right) || E \left( \theta_{n,t|t-1} \right) \right] \]
The KL divergence is defined in terms of topic counts and Dirichlet parameters:

{\small
\begin{gather*}
    E \left( \theta_{n,t|t} \right) = \frac{ A_{n, t-1} \cdot x_{n,t-1|t-1} + \psi_{n, t} + \alpha }{ 1' \left( A_{n, t-1} \cdot x_{n,t-1|t-1} + \psi_{n, t} \right) + K \alpha } \qquad
    E \left( \theta_{n,t|t-1} \right) = \frac{ A_{n, t-1} \cdot x_{n,t-1|t-1} + \alpha }{ 1' A_{n, t-1} \cdot x_{n,t-1|t-1} + K \alpha } \\
    D_{KL} \left[ E \left( \theta_{n,t|t} \right) || E \left( \theta_{n,t|t-1} \right) \right] = \sum_{k=1}^K E \left( \theta_{n,t,k|t} \right) \left[ \log E \left( \theta_{n,t,k|t} \right) - \log E \left( \theta_{n,t,k|t-1} \right) \right]
\end{gather*}
}

We then find the dynamics matrix $A_{n,t-1}$ that minimizes the objective function $\mathcal{L}$ given by Equation \ref{eqn:objective_function},
\begin{gather}
  \label{eqn:objective_function}
  \mathcal{L} = \sum_{t=2}^T \sum_{k=1}^K E \left( \theta_{n,t,k|t} \right) \left[ \log E \left( \theta_{n,t,k|t} \right) - \log E \left( \theta_{n,t,k|t-1} \right) \right]
\end{gather}

Assume that $A_{n,t - 1} \in \mathbb{R}^{K \times K}$ is a diagonal matrix with entries $\mu_{n,t-1,k}$. Taking this into consideration, we can now try to minimize $\mathcal{L}$ by performing gradient descent with respect to the parameters $\mu_{n,t-1,k}$. The gradient term is given in Equation \ref{eqn:gradient_mu},

{\small
\begin{gather}
  \label{eqn:gradient_mu}
  \frac{d \mathcal{L}}{d \mu_{n,t-1,k}} = \frac{d E \left( \theta_{n,t,k|t} \right) }{d \mu_{n,t-1,k}} \left[ \log \frac{ E \left( \theta_{n,t,k|t} \right) } { E \left( \theta_{n,t,k|t-1} \right) } \right] + E \left( \theta_{n,t,k|t} \right) \left[ \frac{d \log E \left( \theta_{n,t,k|t} \right) }{d \mu_{n,t-1,k}} - \frac{d \log E \left( \theta_{n,t,k|t-1} \right) }{d \mu_{n,t-1,k}} \right]
\end{gather}
}
where the individual components can be respectively solved as,
\begin{align}
  \frac{d E \left( \theta_{n,t,k|t} \right) }{d \mu_{n,t-1,k}} &= \frac{ x_{n,t-1,k|t-1} \left( 1' \psi_{n, t} + K \alpha \right) - 1' x_{n,t-1|t-1} \left( \psi_{n, t, k} + \alpha \right) }{ \left[ 1' \left( A_{n, t-1} \cdot x_{n,t-1|t-1} + \psi_{n, t} \right) + K \alpha \right]^2 } \\
  \frac{d \log E \left( \theta_{n,t,k|t} \right) }{d \mu_{n,t-1,k}} &= \frac{1}{E \left( \theta_{n,t,k|t} \right)} \frac{d E \left( \theta_{n,t,k|t} \right) }{d \mu_{n,t-1,k}} \\
  \frac{d \log E \left( \theta_{n,t,k|t-1} \right) }{d \mu_{n,t-1,k}} &= \frac{x_{n,t-1,k|t-1}}{\mu_{n,t-1,k} \cdot x_{n,t-1,k|t-1} + \alpha} - \frac{1' x_{n,t-1|t-1}}{1' A_{n, t-1} \cdot x_{n,t-1|t-1} + K \alpha}
\end{align}

According to Siddiqi et al. \cite{Siddiqi2007}, an LDS is Lyapunov (a.k.a. numerically) stable if the eigenvalues of the dynamics matrix $A_{n,t}$ is less than or equal to one. The eigenvalues of any general matrix are guaranteed to be less than or equals to one if the sum of each row in the matrix is less than or equals to one. To ensure stability of LDTM, we enforce $\mu_{n,t,k}$ to stay within $[0,1]$. By staying within the $[0,1]$ range, the $\mu_{n,t,k}$ is also able to represent decay of the parameters learned in previous time steps.

\subsection{Outline of Parameter Estimation}

Algorithm \ref{chap:gc:alg:model} summarizes the procedure to estimate all parameters of the LDTM model depicted in Figure \ref{chap:gc:fig:plate}. It begins by randomly initializing the latent variables $z_{n, t, m}$, followed by Gibbs sampling iterations which consist of several steps. First, the prior topic distributions are estimated using Kalman Filter. Then the latent variable $z_{n, t, m}$ is sampled by conditioning on the prior parameters $x_{n, t | t - 1}$, sampled variables $\psi_{n, t}$ from previous iterations and the constant parameters $\alpha, \beta$. 
\begin{gather}
  \label{eqn:gibbs_sampling}
  p(z_{n,t,m} = k | x_{n,t|t-1}, \psi_{n,t}, \xi_{k}, \alpha, \beta) = \left( x_{n,t,k|t-1} + \psi_{n,t,k} + \alpha \right) \frac{\xi_{k,m} + \beta}{1' \xi_k + K \beta}
\end{gather}
Using the sampled latent variables, we derive the posterior parameters $x_{n,t | t}$ via Kalman Filter. Finally, we estimate the dynamics matrix $A_{n,t}$ via gradient descent of equation (\ref{eqn:gradient_mu}) for minimizing the KL between the prior and posterior distributions. We repeat these steps until a maximum number of iterations is reached.

\begin{algorithm}[htb]
  \caption{LDTM Inference}
  \label{chap:gc:alg:model}
  \begin{algorithmic}[1]
    \STATE Input: {\small Adoption data for each user $n$ at each time step $t$}
    \STATE Output: Estimated parameters
    \STATE Define: $\psi_{n, t, k}$ is the number of times user $n$ at time $t$ generates topic $k$.
    \STATE Define: $\xi_{k, m}$ is the number of times topic $k$ generates item $m$.
    \STATE // Initialization
    \FOR{$n \leftarrow 1$ \TO $N$}
      \FOR{$t \leftarrow 1$ \TO $T_n$}
        \FOR{$m \leftarrow 1$ \TO $M_{n, t}$}
          \STATE $k \leftarrow uniformRandom(1, K)$ \;
          \STATE $\psi_{n, t, k} \leftarrow \psi_{n, t, k} + 1$; 
          \STATE $\xi_{k, m} \leftarrow \xi_{k, m} + 1$; 
          \STATE $z_{n, t, m} \leftarrow k$;
        \ENDFOR
      \ENDFOR
    \ENDFOR
    \STATE // Gibbs sampling iterations
    \REPEAT
      \FOR{$n \leftarrow 1$ \TO $N$}
        \STATE // Estimating topic distribution parameters: Kalman filter
        \FOR{$t \leftarrow 1$ \TO $T_n$}
          \STATE $x_{n, t | t - 1} \leftarrow A_{n, t - 1} \cdot x_{n, t - 1 | t - 1}$;
          \FOR{$m \leftarrow 1$ \TO $M_{n, t}$}
            \STATE $k \leftarrow z_{n, t, m}$; 
            \STATE $\psi_{n, t, k} \leftarrow \psi_{n, t, k} - 1$; 
            \STATE $\xi_{k, m} \leftarrow \xi_{k, m} - 1$;
            \STATE $k \leftarrow sample(m, x_{n, t | t - 1} + \psi_{n, t}, \xi_{k}, \alpha, \beta)$;    // Sample using Equation (\ref{eqn:gibbs_sampling}).
            \STATE $\psi_{n, t, k} \leftarrow \psi_{n, t, k} + 1$; 
            \STATE $\xi_{k, m} \leftarrow \xi_{k, m} + 1$; 
            \STATE $z_{n, t, m} \leftarrow k$;
          \ENDFOR
          \STATE $x_{n, t | t} \leftarrow x_{n, t | t-1} + \psi_{n, t}$;
        \ENDFOR
        \STATE // Estimating decay parameters: KL divergence minimization 
        \FOR{$t \leftarrow 2$ \TO $T_n$}
          \FOR{$k \leftarrow 1$ \TO $K$}
            \STATE Update the diagonal entries $\mu_{n,t-1,k} \in A_{n, t-1}$ using Equation (\ref{eqn:gradient_mu})
          \ENDFOR
        \ENDFOR
      \ENDFOR
    \UNTIL{maximum iterations}
  \end{algorithmic}
\end{algorithm}

\section{Computing Temporal Social Correlation in Topic Space using Granger Causality}
\label{chap:gc:sec:granger_causality}


After obtaining the posterior topic distributions $\theta_{n, t | t}, \forall n \in U$, we can construct time series of the distributions and calculate the Temporal Social Correlation (TSC) using \emph{Granger causality} (GC) \cite{Granger1980}. For a pair of users ($i, j$), $TSC$ can be measured in two directions, $TSC( i \rightarrow j, \tau )$ and $TSC( j \rightarrow i, \tau )$, pivoted at a specific time step $\tau$. One should appropriately choose $\tau$ to indicate the starting point for information transfer between $i$ and $j$. Given $\tau$, we could then select a time window $[\tau - W, \tau + L]$ to constrain time series used for comparison, where $L$ is the number of time steps to ``lookahead'' for measuring $TSC$ and $W$ is the ``width'' of past time steps for predicting the future.

For notational simplicity, we denote the topic distributions for users $i$ and $j$ at $t$ as $i_t$ and $j_t$ respectively. Specifically, given two users $i$ and $j$ who interact at time $\tau$, $TSC(i \rightarrow j, \tau)$ is computed as follows:
\begin{enumerate}
   \item Formulate the two linear regression tasks:
  \begin{gather}
    \tilde{j}_t = \eta_0 + \left( \sum_{w = 1}^{W} \eta_w j_{t - w} \right) \nonumber \\ 
    \label{chap:gc:eqn:gc1}
    R_1 = \sum_{t = \tau}^{\tau + L} \left( j_t - \tilde{j}_t \right)' \left( j_t - \tilde{j}_t \right)
  \end{gather}
  \begin{gather}
    \bar{j}_t = \eta_0 + \left( \sum_{w = 1}^{W} \eta_w j_{t - w} + \lambda_w i_{t - w} \right) \nonumber \\ 
    \label{chap:gc:eqn:gc2}
    R_2 = \sum_{t = \tau}^{\tau + L} \left( j_t - \bar{j}_t \right)' \left( j_t - \bar{j}_t \right)
  \end{gather}
  where $\tau$ is the time point when $i$ and $j$ begins transferring information between one another. 

  \item Estimate for the parameters $\left \{ \eta_0, \ldots, \eta_W \right \}$ by minimizing the least squares error in (\ref{chap:gc:eqn:gc1}) using \emph{coordinate descent} \cite{Bezdek1987}, and then estimate \emph{only} for the parameters $\left \{ \lambda_1, \ldots, \lambda_W \right \}$ by minimizing (\ref{chap:gc:eqn:gc2}). The first linear regression given by (\ref{chap:gc:eqn:gc1}) uses $j$'s past information to predict $j$'s future, while the second linear regression (\ref{chap:gc:eqn:gc2}) uses additional information from $i$'s past to predict $y$'s future.

  \item To obtain the $TSC(i \rightarrow j, \tau)$, we measure how much $i$'s past improves the prediction of $j$'s future by computing the F-statistic (\textit{F-stat}),
  \[ TSC(i \rightarrow j, \tau) = \textit{F-stat} = \frac{R_1 - R_2}{R_2} \cdot \frac{2L - 1}{W} \]
  Because the formula (\ref{chap:gc:eqn:gc2}) uses more parameters than (\ref{chap:gc:eqn:gc1}), the sum-of-squares error given by $R_2$ is always smaller than $R_1$, i.e. $R_2 < R_1$, which implies that \textit{F-stat} is always positive.

  \item Repeat the steps for computing $TSC( j \rightarrow i, \tau )$ and compare whether $TSC( i \rightarrow j, \tau ) > TSC( j \rightarrow i, \tau )$ or otherwise. 
\end{enumerate}

\section{Experiments}
\label{chap:gc:sec:experiments}

To evaluate the effectiveness of LDTM and the $TSC$ calculated for pairs of users, we require datasets that provide users' temporal adoptions and the interactions between users that lead to information transfer between them. The publicly available DBLP \cite{Ley2005} and ACM Digital Library (ACMDL) \cite{ACM} bibliographic datasets provide the information we require. We first describe how we obtain subsets of the data from DBLP and ACMDL for our evaluation needs. Then we evaluate the effectiveness of LDTM for several scenarios of the dynamics matrix $A_{n,t}$:
\begin{enumerate}
  \item LDA: To reduce LDTM to the baseline LDA, we simply set $A_{n,t}$ as identity matrix for every user $n$ and every time step $t$, i.e. $A_{n,t} = I$.
  \item Half Decay: We set $A_{n,t}$ as diagonal matrix with constant values of 0.5, i.e. $A_{n,t} = 0.5 \cdot I$.
  \item Full Decay: We set $A_{n,t}$ as zero matrix, i.e. $A_{n,t} = 0$.
  \item LDTM: We automatically determine the values of the dynamics matrix $A_{n,t}$.
\end{enumerate}

We show that automatically estimating $A_{n,t}$ in the LDTM case gives us better representations of authors' temporal adoption behavior than setting constant values for $A_{n,t}$. Using a case study as example, we show how the topic distributions over time for an author and his co-authors can be used to determine the information transfer relationship between them. Finally, we compare and compute the $TSC$ between every pair of authors using topic distributions from the four scenarios of the dynamics matrix $A_{n,t}$ (i.e., LDA, Half Decay, Full Decay and LDTM).



\subsection{Data Set}

We used the DBLP and ACMDL data to obtain the required users and items. The authors who wrote papers together are treated as users, and the words in their papers are seen as adopted items. We used the words in the abstract for ACMDL, and those in the paper title for DBLP. The co-authorship information provides a time point where interaction occurred between the two authors.

Given the large number of publications in DBLP and ACMDL, we only used a subset of papers from DBLP and ACMDL. We sampled a data subset that covers a wide variety of fields in computer science, with the papers published in the Journal of ACM (JACM) as a seed set. We then expanded the coverage by including other non-JACM publications by authors with at least one JACM publication. The sample set obtained here is termed \emph{ego-1}. By including the co-authors of the authors in \emph{ego-1} and their papers, we get a larger set called \emph{ego-2}. We repeat the process once again to get \emph{ego-3}.


\begin{table}[htb]
  \caption{Data Set Sizes}
  \label{chap:gc:table:dataset_size}
  \begin{tabular}{|c|r|r|r|}
    \hline
                  & \#authors & \#words & period    \\
    \hline
    ACMDL (ego-2) &    24,569 &  33,044 & 1952-2011 \\
    \hline
    ACMDL (ego-3) &   157,715 &  44,308 & 1952-2011 \\
    \hline
    DBLP (ego-2)  &    52,754 &  20,080 & 1936-2013 \\
    \hline
    DBLP (ego-3)  &   388,092 &  40,463 & 1936-2013 \\
    \hline
  \end{tabular}
\end{table}

Table \ref{chap:gc:table:dataset_size} gives the sizes of the \emph{ego-2} and \emph{ego-3} datasets. DBLP has more authors than ACMDL, because DBLP covers a longer history of publications and has more sources of publications. On the other hand, ACMDL focuses mainly on ACM-related publications. After pruning the stop-words and non-frequent (less than ten occurrences) words, the ACMDL sampled dataset have slightly more words than DBLP, as ACMDL provides words in the abstract of publications while DBLP only has words in the paper titles. We used the smaller \emph{ego-2} samples for experiments that require repetitions, and the much larger \emph{ego-3} samples for experiments that only require a single run. Because JACM lists a total of 26 major fields in Computer Science, we used 26 as the number of topics for training our models in all the subsequent experiments.

\subsection{Convergence of Log Likelihood}

We first evaluate the convergence of the log likelihood for the case where the dynamics matrix $A_{n,t}$ is automatically computed, and for other cases where $A_{n,t}$ is set to constant values. We used the \emph{ego-2} samples for evaluating log likelihood convergence, because \emph{ego-2} will be used later for the predictive evaluations.

\begin{figure}[htb]
  \centering
  \subfigure[ACMDL]
  {
    \includegraphics[width=2.2in]{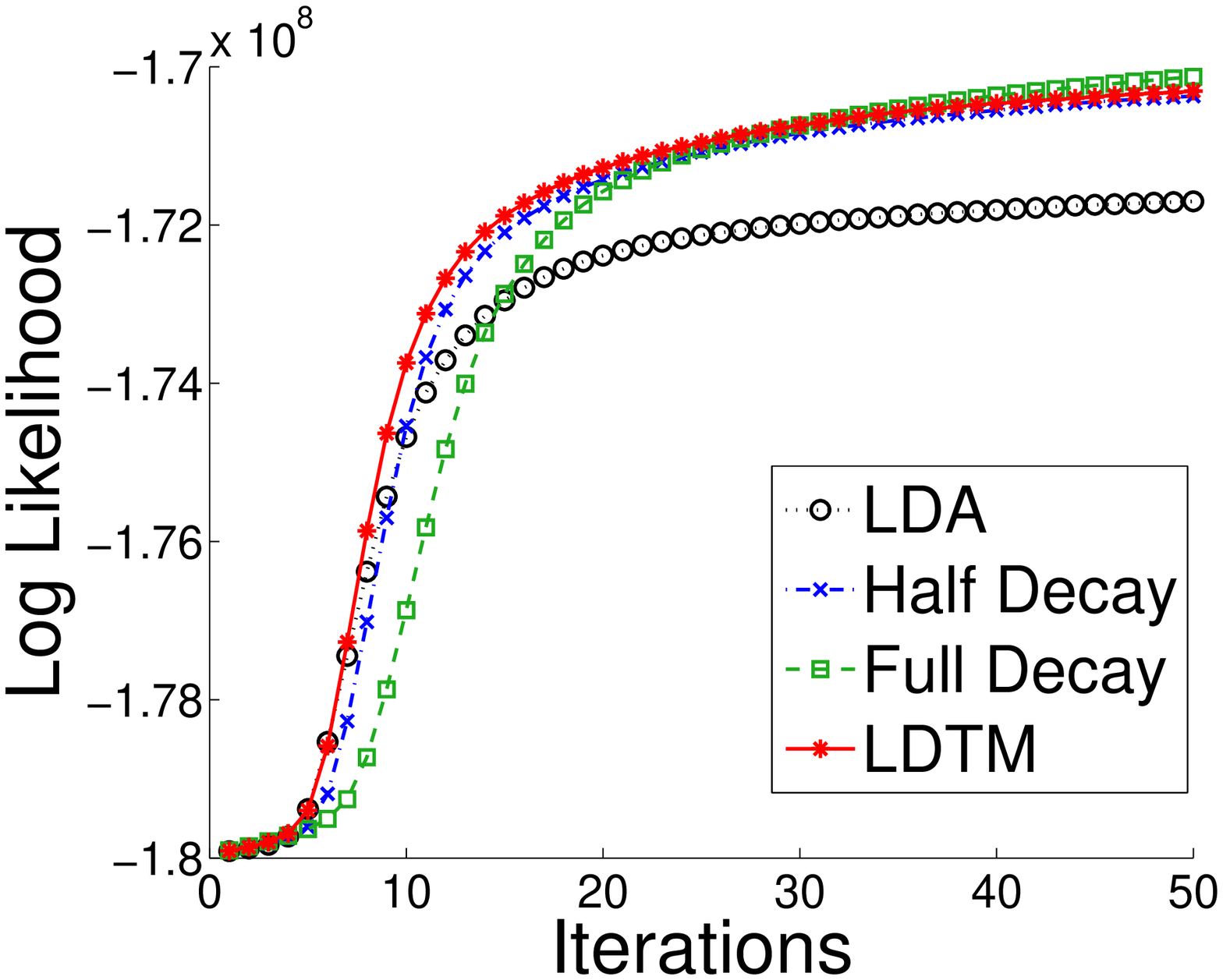}
    \label{fig:acmdl_logll_vs_iterations}
  }
  \subfigure[DBLP]
  {
    \includegraphics[width=2.2in]{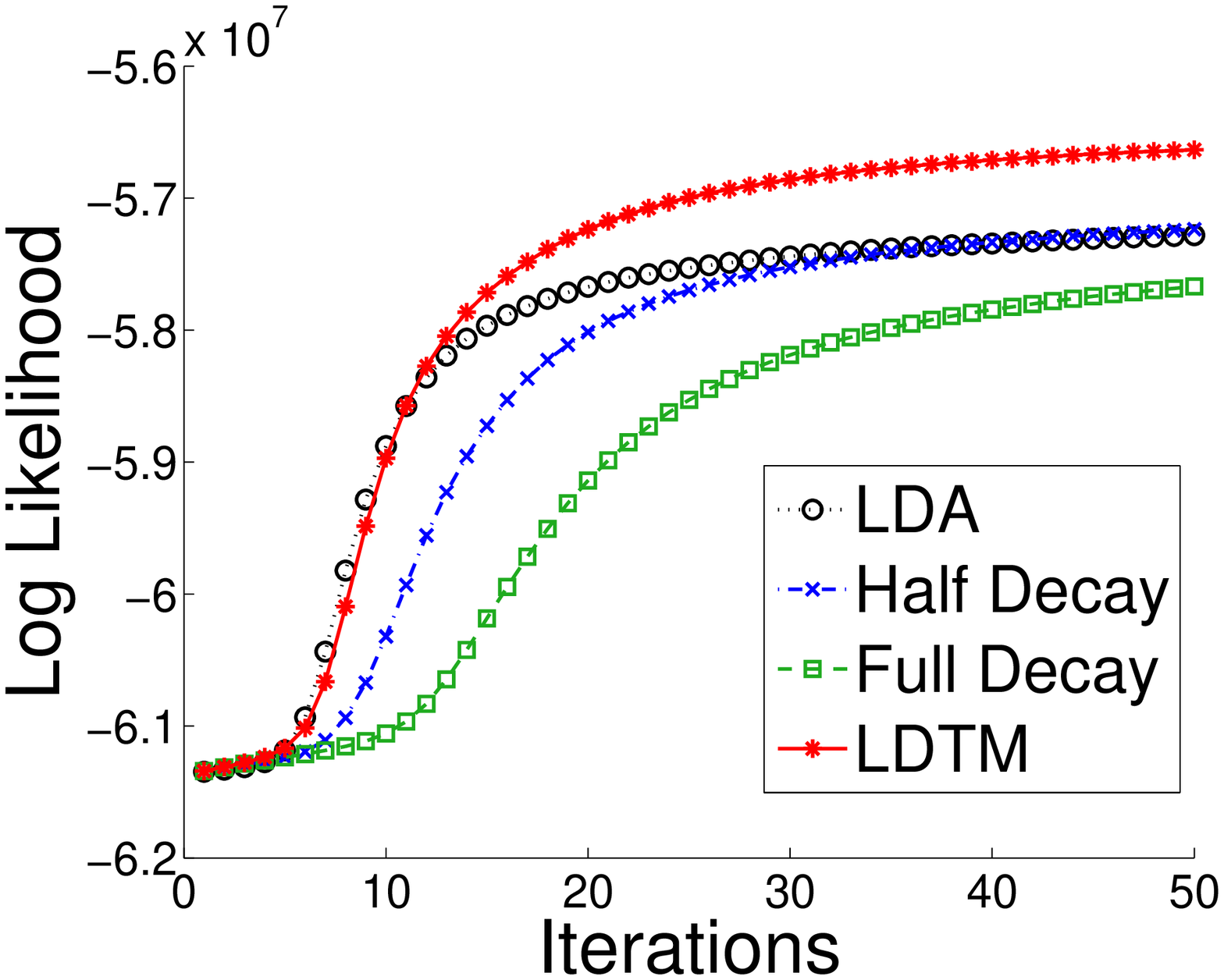}
    \label{fig:dblp_logll_vs_iterations}
  }
  \caption{Log Likelihood vs \# of Iterations}
  \label{fig:logll_vs_iterations}
\end{figure}

Figures \ref{fig:acmdl_logll_vs_iterations} and \ref{fig:dblp_logll_vs_iterations} show how the log likelihood varies with the number of iterations for ACMDL (ego-2) and DBLP (ego-2). We can see that LDTM achieves the highest likelihood in DBLP and is able to converge with log likelihood comparable to that of Half Decay and Full Decay.

Figures \ref{fig:acmdl_logll_vs_iterations} and \ref{fig:dblp_logll_vs_iterations} also reveal another interesting observation: the Full Decay model is able to perform well in ACMDL, but not as well in DBLP. This can be explained as follows, since ACMDL provide words from papers' abstract, the information within each time step is sufficient for estimating accurate parameters. But in DBLP, only the words in the paper titles are available, providing less information for parameter estimation. As a result, decaying the parameters of previous time steps does not allow Full Decay to leverage on the previously observed data, which explains the poor likelihood in DBLP. The same explanation applies to LDA, since LDA shows an opposite performance to Full Decay.

Based on the ACMDL and DBLP results in Figure \ref{fig:logll_vs_iterations}, we can see that fixing the decay parameters of the dynamics matrix does not give consistent performance in comparison to that of LDTM. This shows that the automatic estimation of dynamics matrix in LDTM can better model the different properties of the available data.





\subsection{Comparison Results on Held-out Test Set}

We compared the automatic estimation of dynamics matrix $A_{n, t}$ for LDTM against the fixed values of $A_{n, t}$ (i.e., Full Decay, Half Decay, LDA) in a prediction task. We repeated the prediction experiments for five runs and took the average results. For each run, we generated five sets of training and testing data by hiding in incremental proportions of $10\%$ from the sampled ACMDL (ego-2) and DBLP (ego-2) datasets. When creating the test sets, we ensure that each subsequent test set is a superset of the previous test set. We trained LDTM and other baseline models on the remaining data sets for 50 iterations, and evaluated their predictive performances on the held-out test sets. The predictive performance on the held-out test sets is measured in terms of \emph{average log likelihood} for each time step $t$ ($ALL@t$), defined as, 
\[ ALL@t = \frac{ \sum_{n}^N \sum_{m}^{M_{n,t}} \log p(w_{n,t,m} | \theta_n, \phi) }{ \sum_{n}^N M_{n,t} } \]
Essentially, the $ALL@t$ gives us a measure of how well the estimated parameters can predict the test sets. Normalization of the log likelihood (over the total number of words) in each time step is necessary to avoid over-deflating the log likelihood at different time steps. For example, larger $t$ would have more words and hence smaller (more negative) log likelihood as compared to smaller $t$. The log likelihood normalization at each $t$ would thus allow for a better comparison across different time steps. A higher $ALL@t$ suggests a better predictive performance for the respective model.

\begin{figure}[htb]
  \centering
  \subfigure[Test size=20\%]
  {
    \includegraphics[width=2.2in]{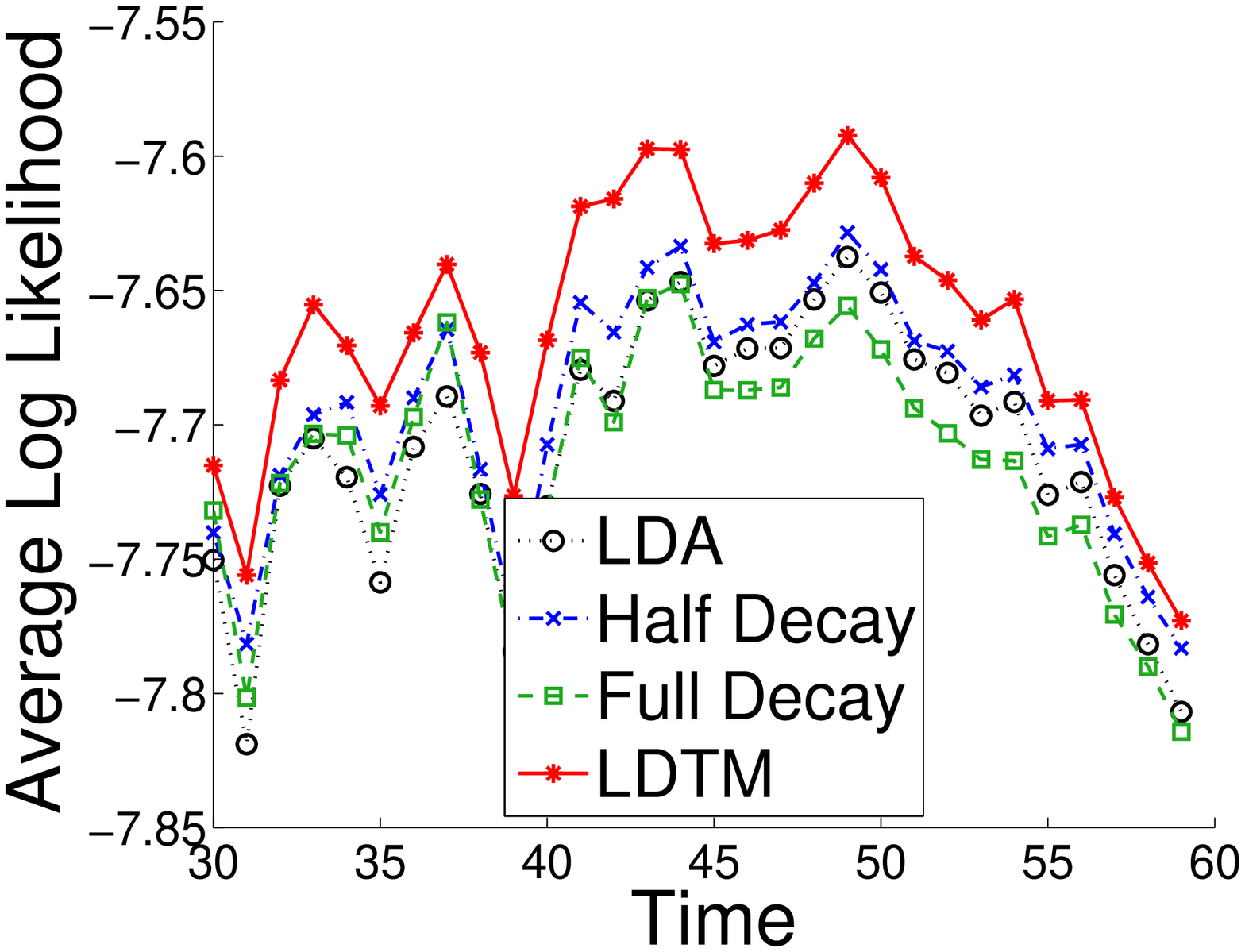}
    \label{fig:acmdl_avg_logll_over_time_2}
  }
  \subfigure[Test size=30\%]
  {
    \includegraphics[width=2.2in]{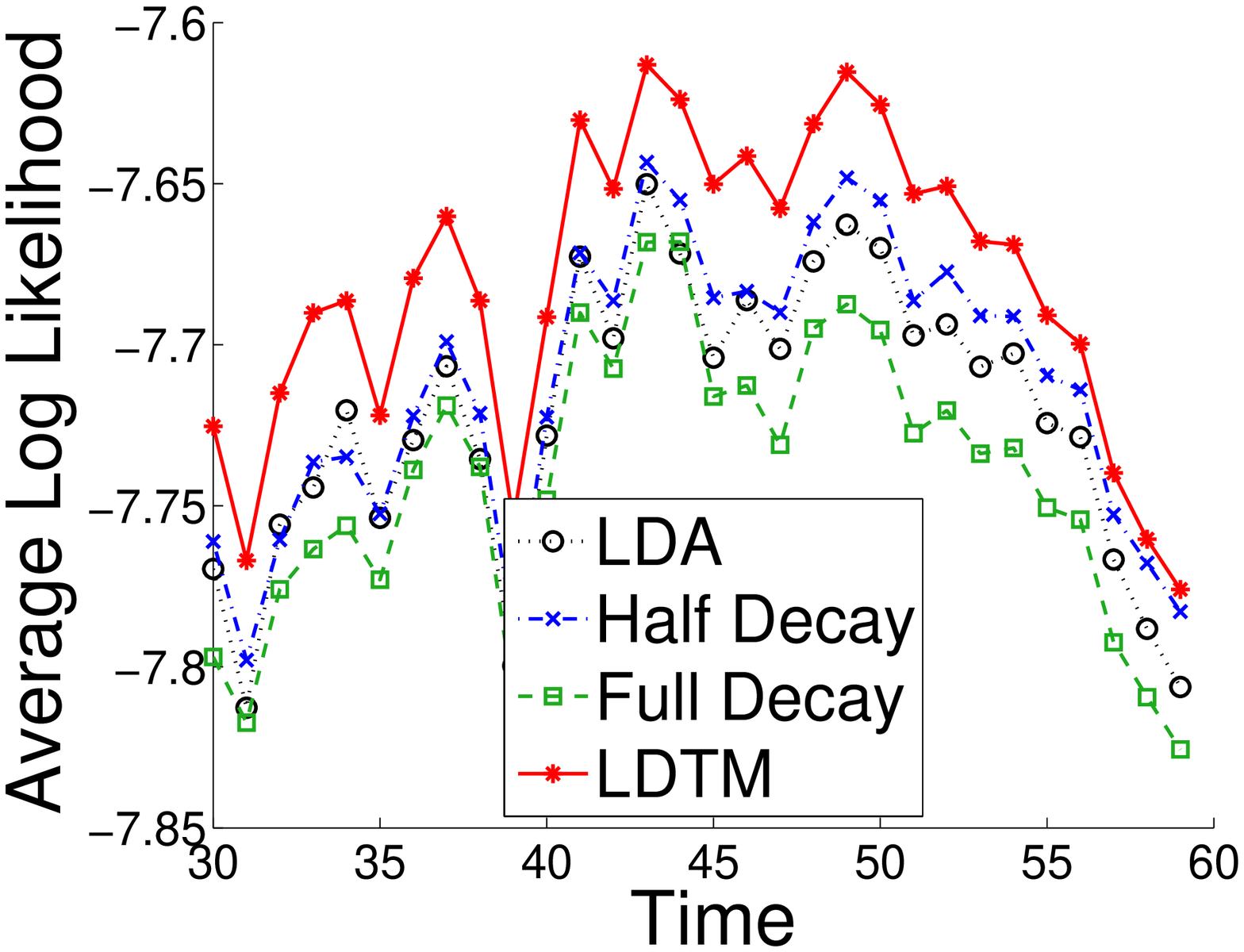}
    \label{fig:acmdl_avg_logll_over_time_3}
  }
  \subfigure[Test size=40\%]
  {
    \includegraphics[width=2.2in]{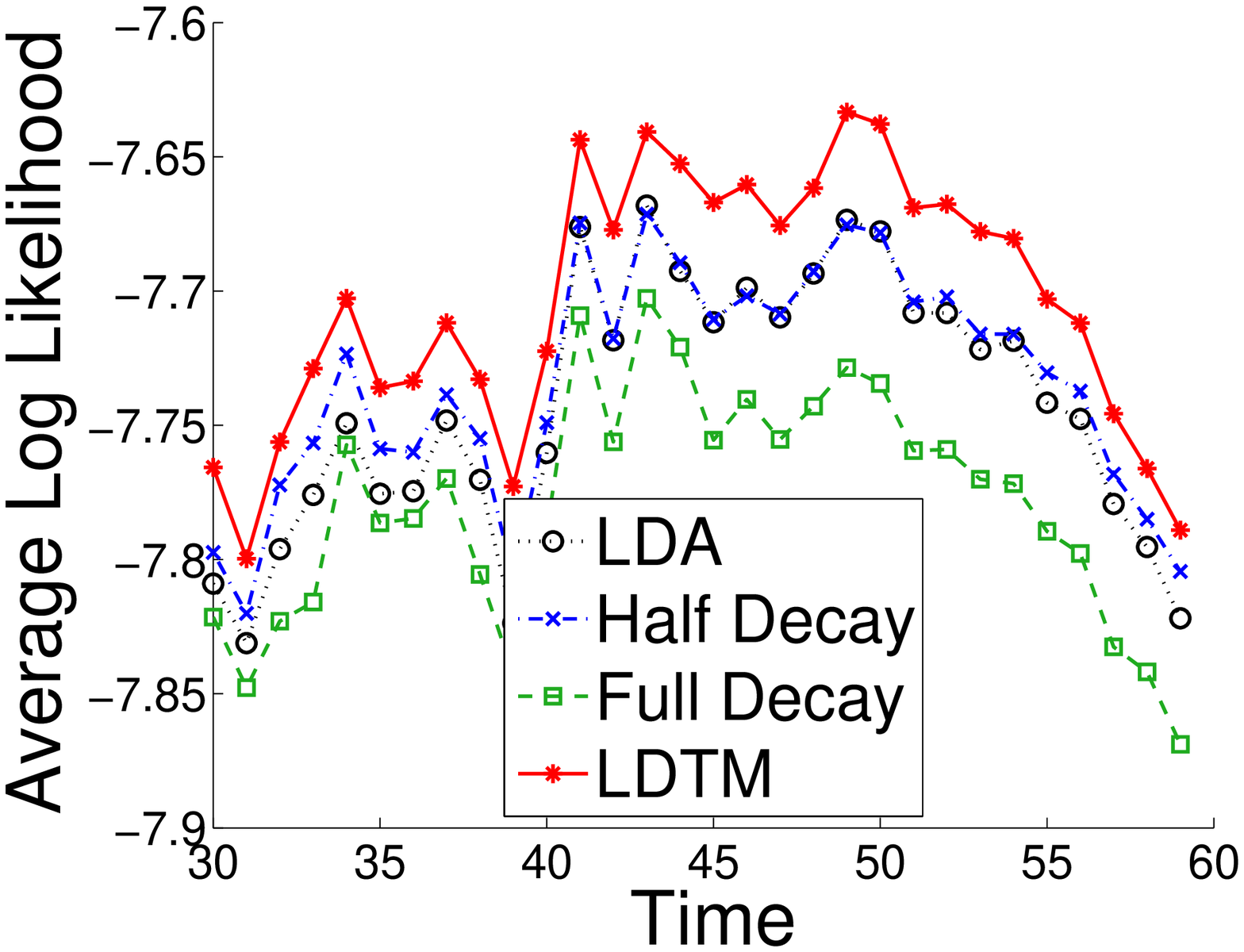}
    \label{fig:acmdl_avg_logll_over_time_4}
  }
  \subfigure[Test size=50\%]
  {
    \includegraphics[width=2.2in]{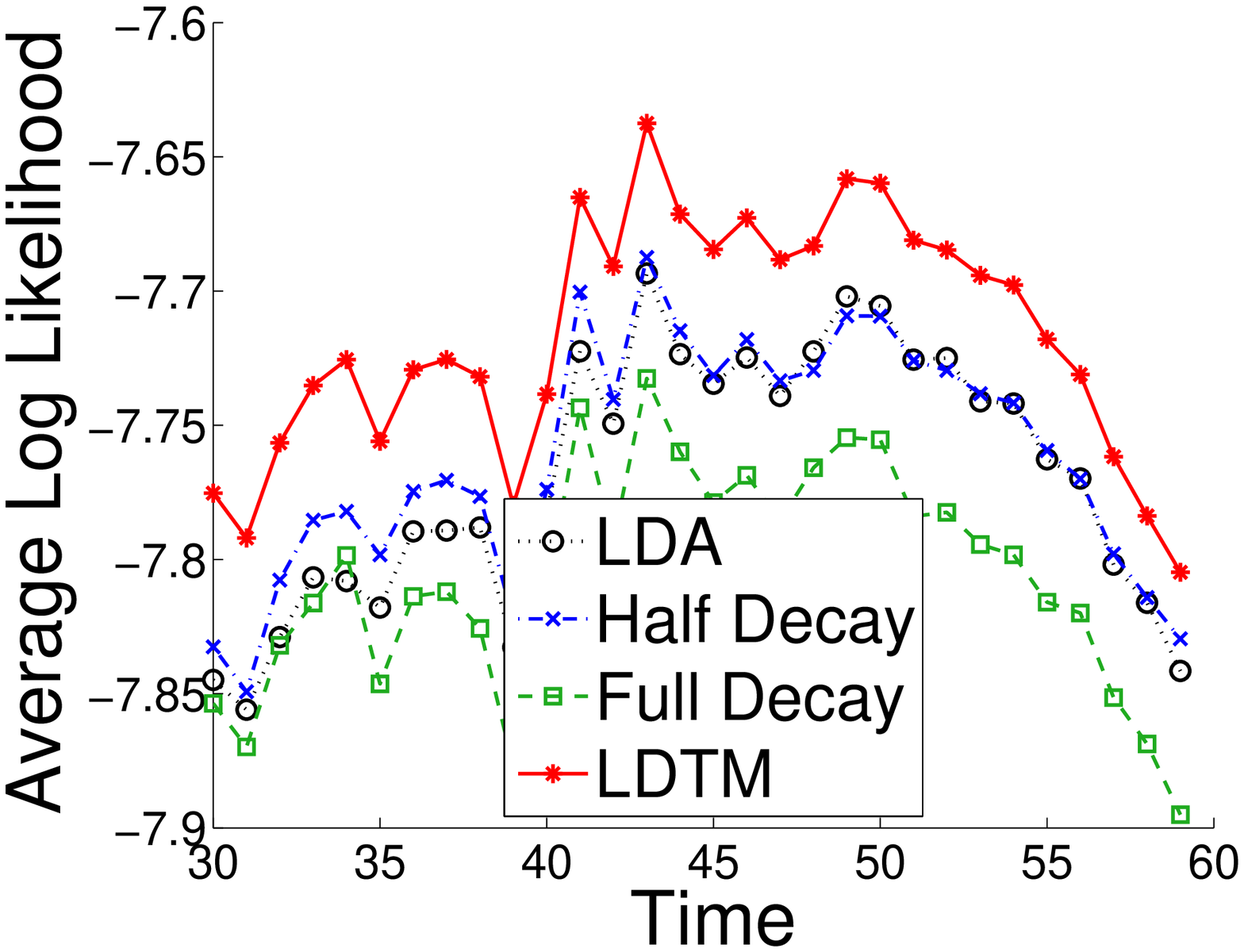}
    \label{fig:acmdl_avg_logll_over_time_5}
  }
  \caption{ACMDL: Log Likelihood over Time for Held-out Test Set}
  \label{fig:acmdl_avg_logll_over_time}
\end{figure}

\begin{figure}[htb]
  \centering
  \subfigure[Test size=20\%]
  {
    \includegraphics[width=2.2in]{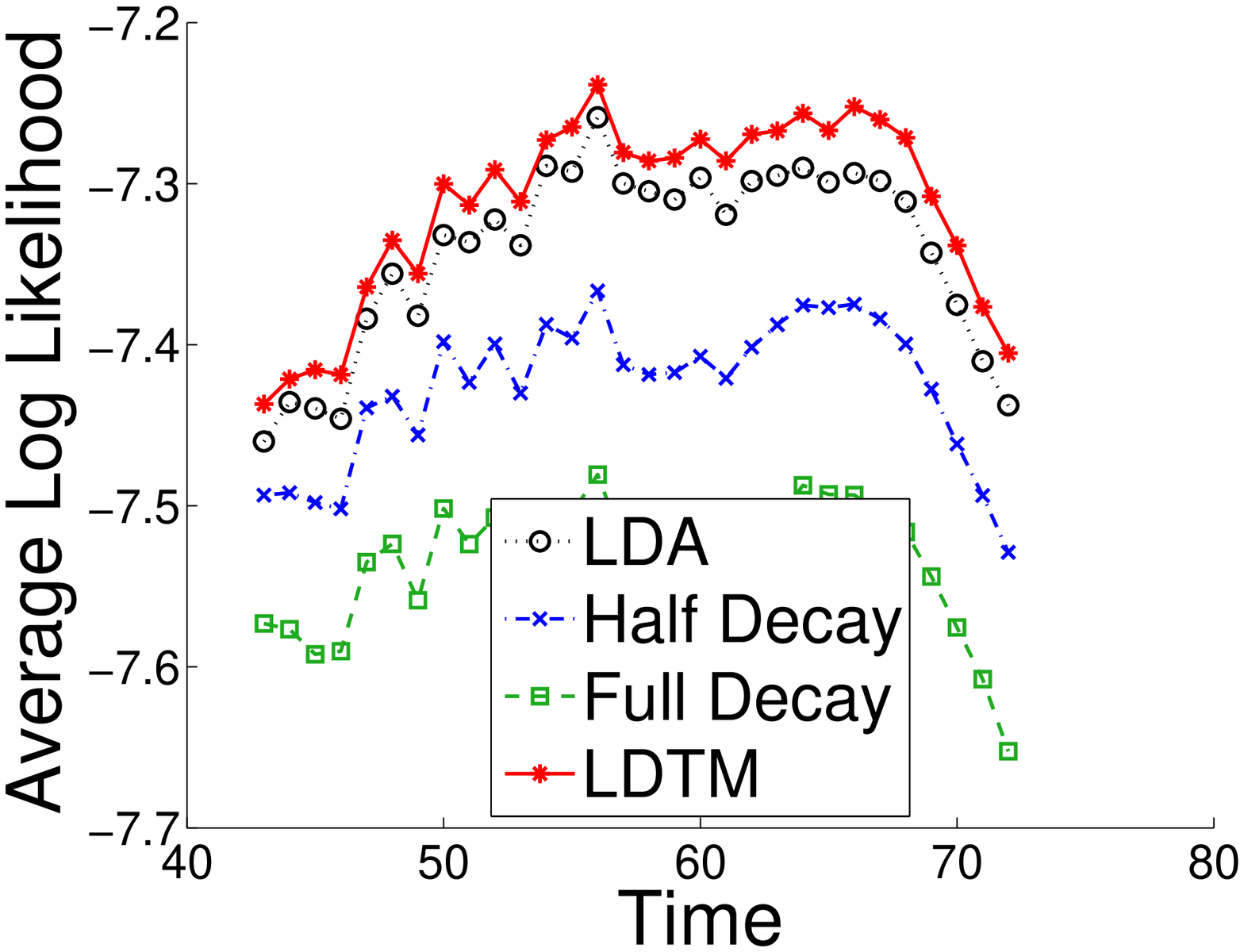}
    \label{fig:dblp_avg_logll_over_time_2}
  }
  \subfigure[Test size=30\%]
  {
    \includegraphics[width=2.2in]{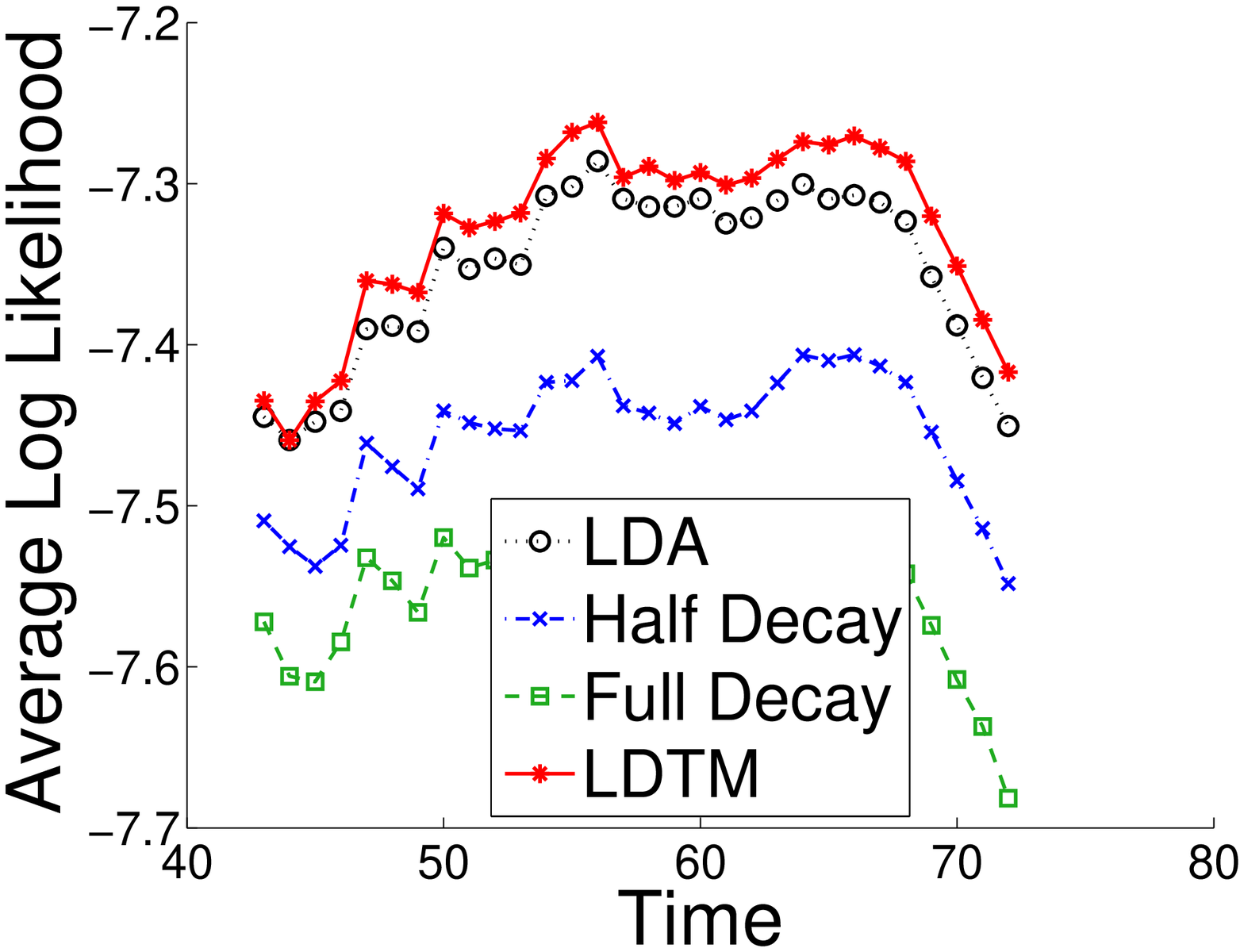}
    \label{fig:dblp_avg_logll_over_time_3}
  }
  \subfigure[Test size=40\%]
  {
    \includegraphics[width=2.2in]{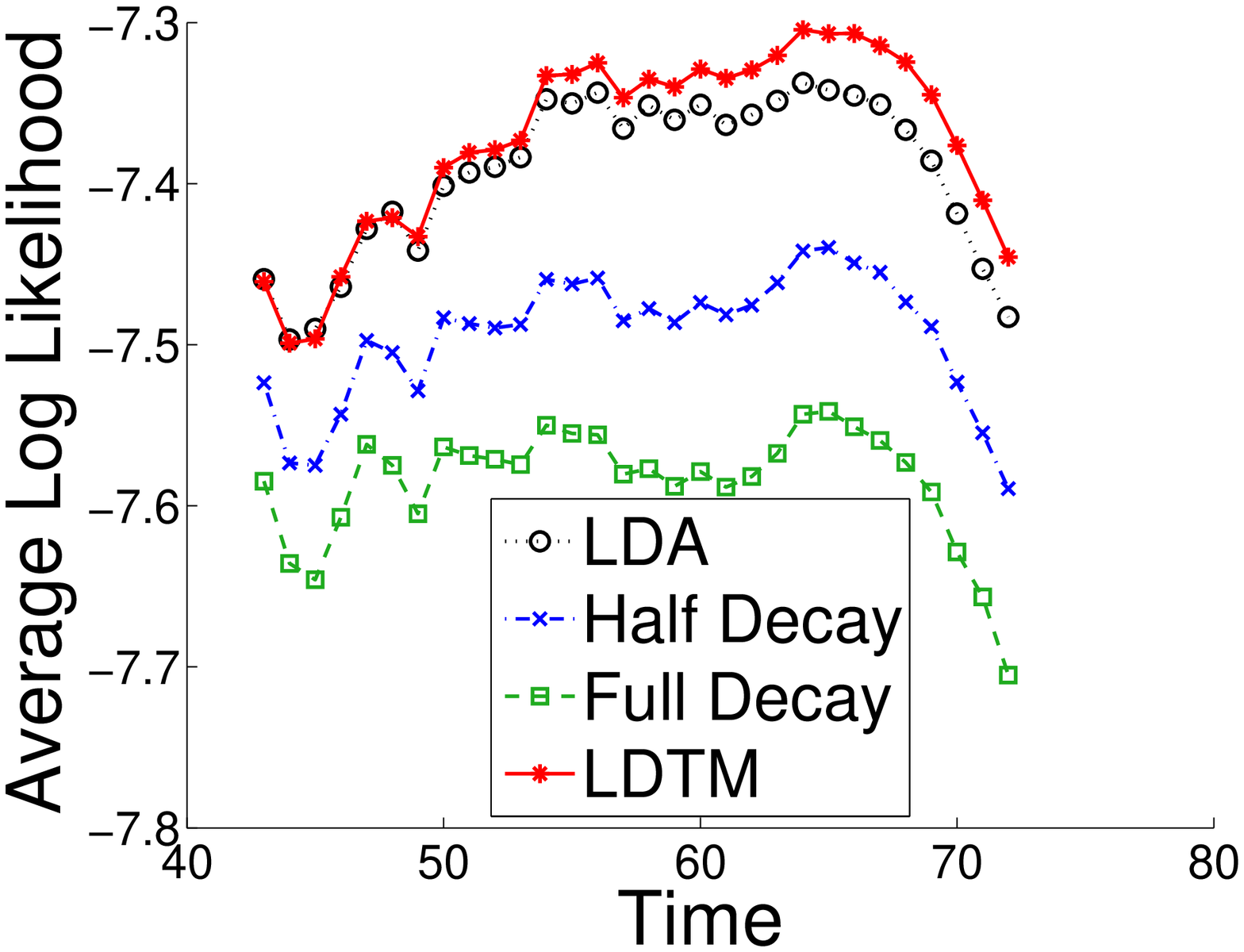}
    \label{fig:dblp_avg_logll_over_time_4}
  }
  \subfigure[Test size=50\%]
  {
    \includegraphics[width=2.2in]{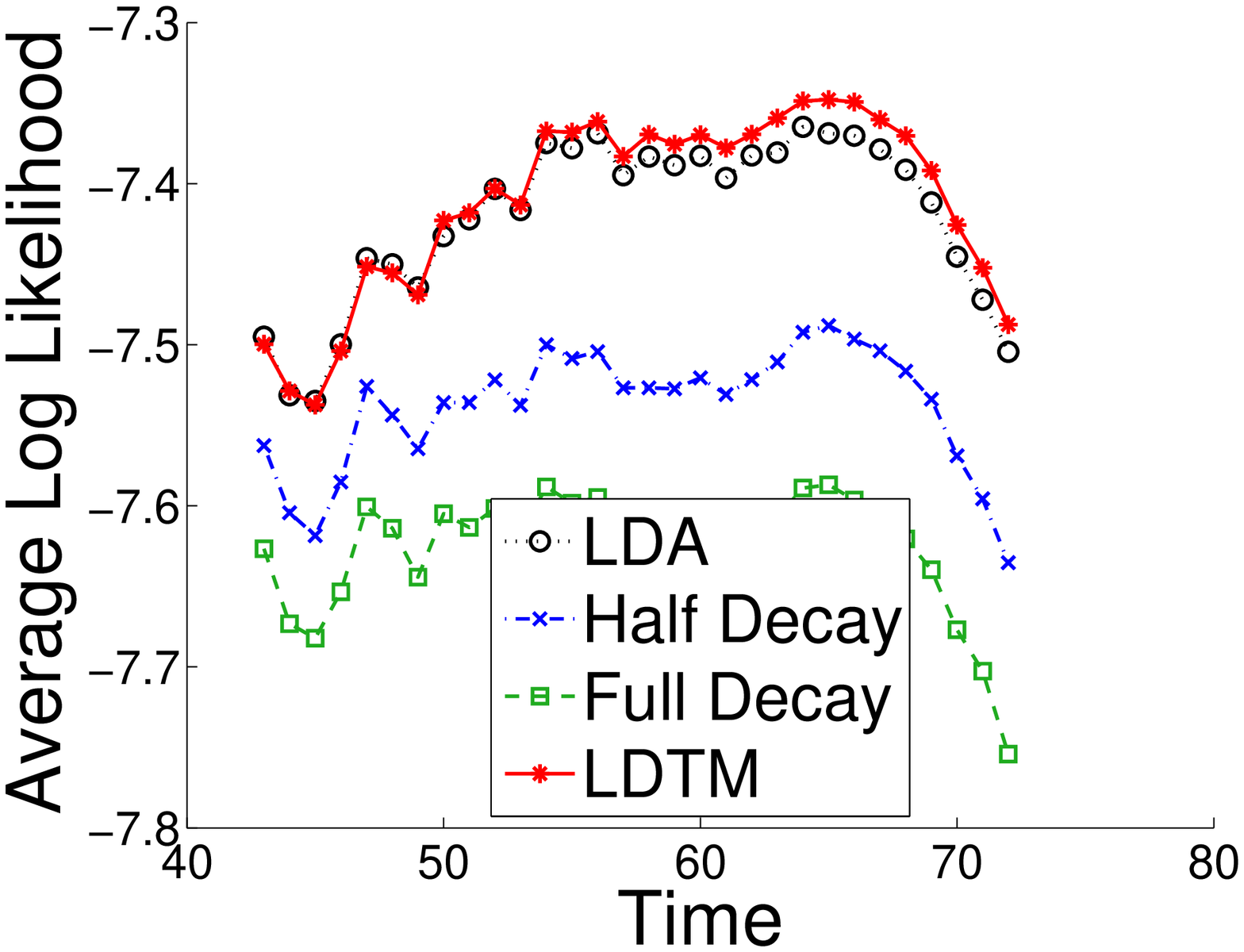}
    \label{fig:dblp_avg_logll_over_time_5}
  }
  \caption{DBLP: Log Likelihood over Time for Held-out Test Set}
  \label{fig:dblp_avg_logll_over_time}
\end{figure}

Figures \ref{fig:acmdl_avg_logll_over_time_2}-\ref{fig:acmdl_avg_logll_over_time_5} and \ref{fig:dblp_avg_logll_over_time_2}-\ref{fig:dblp_avg_logll_over_time_5} show the $ALL@t$ for every $t$. All these figures show that LDTM outperforms all other baseline models in terms of $ALL@t$, while Full Decay performs the worst. Although Full Decay gives highest log likelihood performance in the convergence results as shown in Figure \ref{fig:acmdl_logll_vs_iterations}, the result in Figure \ref{fig:acmdl_avg_logll_over_time} does not show the same correlations, which may be attributed to the fact that Full Decay overfits the data. Meanwhile, Half Decay shows that it fits the data relatively well (cf. Figure \ref{fig:logll_vs_iterations}), but it does not perform as well as LDTM (cf. Figure \ref{fig:acmdl_avg_logll_over_time}) and it performs worse than LDA (cf. Figure \ref{fig:dblp_avg_logll_over_time}). While LDA performs better than Half Decay and Full Decay in Figure \ref{fig:dblp_avg_logll_over_time}, it does not outperform Half Decay in Figure \ref{fig:acmdl_avg_logll_over_time}. By contrast, LDTM is able to achieve consistently good performance in both datasets, winning over Full Decay, Half Decay and LDA. This suggests that the dynamics matrix estimated using the algorithm in Section \ref{sec:estimate_dynamics} can capture the dynamic patterns of each user more accurately.

\subsection{Case Study}

This section provides a case study to illustrate Granger causality among authors. We first show how an author's topic interests change over the years. Subsequently, we describe how the changes in his co-authors' topic interests explain the changes in his own topic interests. 

In this study, we focus on the profile of Professor Duminda Wijesekera (D. Wijesekera), so as to remain consistent with the case study in our earlier work \cite{Chua2012}. For this case study, we performed our analysis on the DBLP (ego-3) dataset, due to its wider coverage of authors and years. Based on our earlier results of various temporal topic models on the DBLP dataset, we choose to analyze the Granger causality based on the topic distributions computed by LDTM and LDA, as the two yielded the best performance on DBLP. By comparing LDTM and LDA, we also show the importance of using the correct latent factor model for Granger causality analysis.

\subsubsection{Granger Causality using LDTM}

\begin{figure}[!htb]
  \centering
  \subfigure[Top Four Topics]
  {
    \includegraphics[width=2.2in]{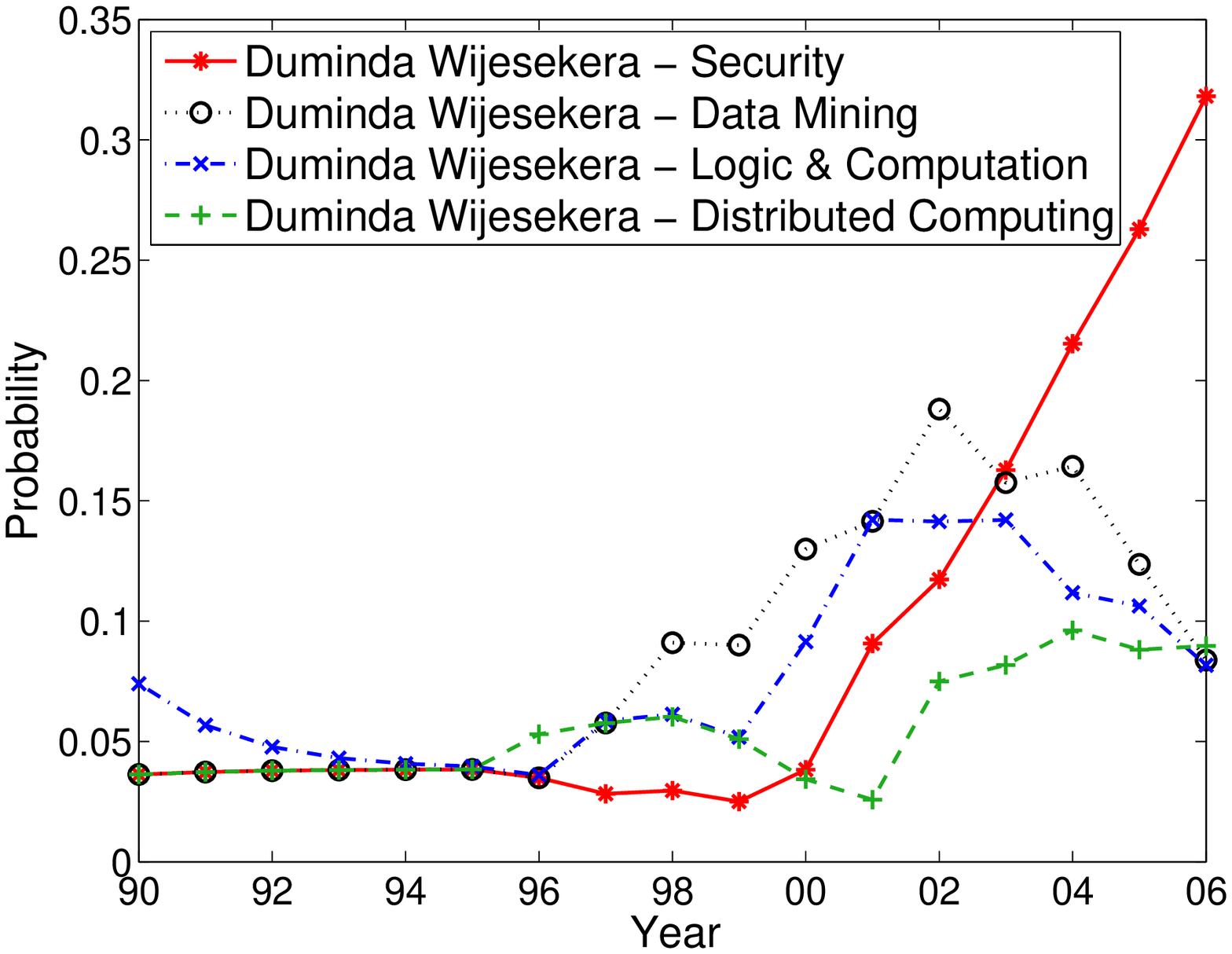}
    \label{fig:duminda_wijesekera_lds_topics}
  }
  \subfigure[Co-Authors' Security Topic Proportion]
  {
    \includegraphics[width=2.2in]{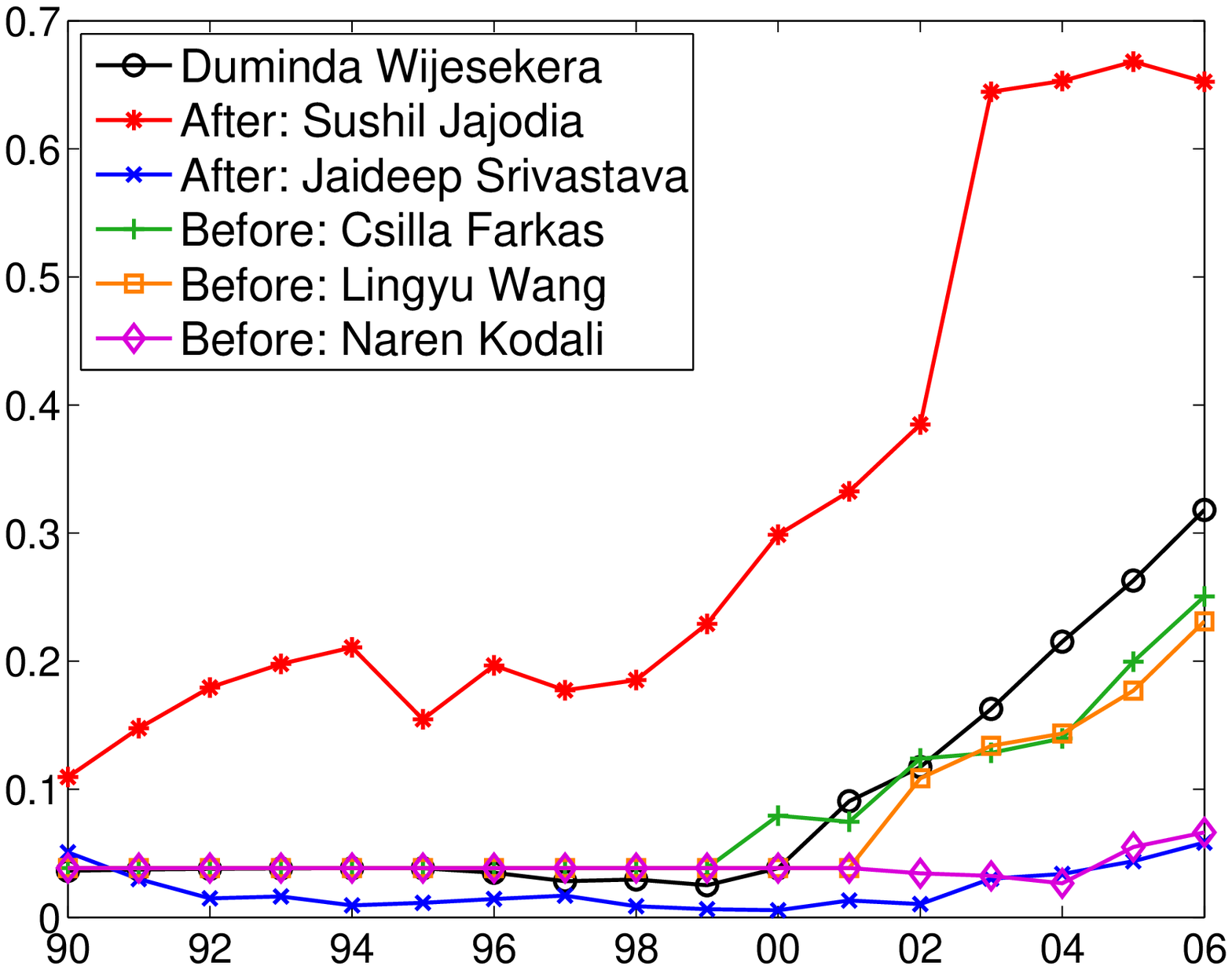}
    \label{fig:duminda_wijesekera_coauthors_lds_topics}
  }
  \caption{LDTM Results: Duminda Wijesekera's and His Co-Authors' Topic Interests from 1990 to 2006}
  \label{fig:duminda_and_coauthor_lds_topics}
\end{figure}

\begin{table}[!htb]
  \caption{Topics Derived from LDTM}
    \label{chap:gc:tbl:ldtm_topics}
    \begin{tabular} {|c|c|c|c|}
      \hline
      \textbf{Security} & \textbf{Data Mining} & \textbf{Logic \& Computation} & \textbf{Distributed Computing} \\
      \hline
      security & data & logic & service \\
      secure & mining & verification & management \\
      scheme & fuzzy & programming & grid \\
      efficient & databases & reasoning & computing \\
      privacy & query & languages & framework \\
      \hline
    \end{tabular}    
\end{table}

Figure \ref{fig:duminda_wijesekera_lds_topics} shows the LDTM-induced topic distribution of D. Wijesekera from year 1990 to 2006 (top four topics), with the corresponding topic words shown in Table \ref{chap:gc:tbl:ldtm_topics}. During this period, the ``Security'' topic has the largest area under the curve. For illustration purposes, we show the ``Security'' topic proportion of D. Wijesekera's co-authors for the same time period. 

Due to space limitation, we only show five co-authors who collaborated with D. Wijesekera frequently in Figure \ref{fig:duminda_wijesekera_coauthors_lds_topics}. Among them, Sushil Jajodia's and Jaideep Srivastava's names are placed \emph{after} D. Wijesekera in the papers they wrote, while Csilla Farkas', Lingyu Wang's and Naren Kodali's names appear \emph{before} D. Wijesekera. From Figure \ref{fig:duminda_wijesekera_coauthors_lds_topics}, we can also see clearly that D. Wijesekera's \textbf{-o-} line follows the trend of Sushil Jajodia's \textcolor{red}{-*-} line. 

\subsubsection{Granger causality using LDA}

\begin{figure}[!htb]
  \centering
  \subfigure[Top Four Topics]
  {
    \includegraphics[width=2.2in]{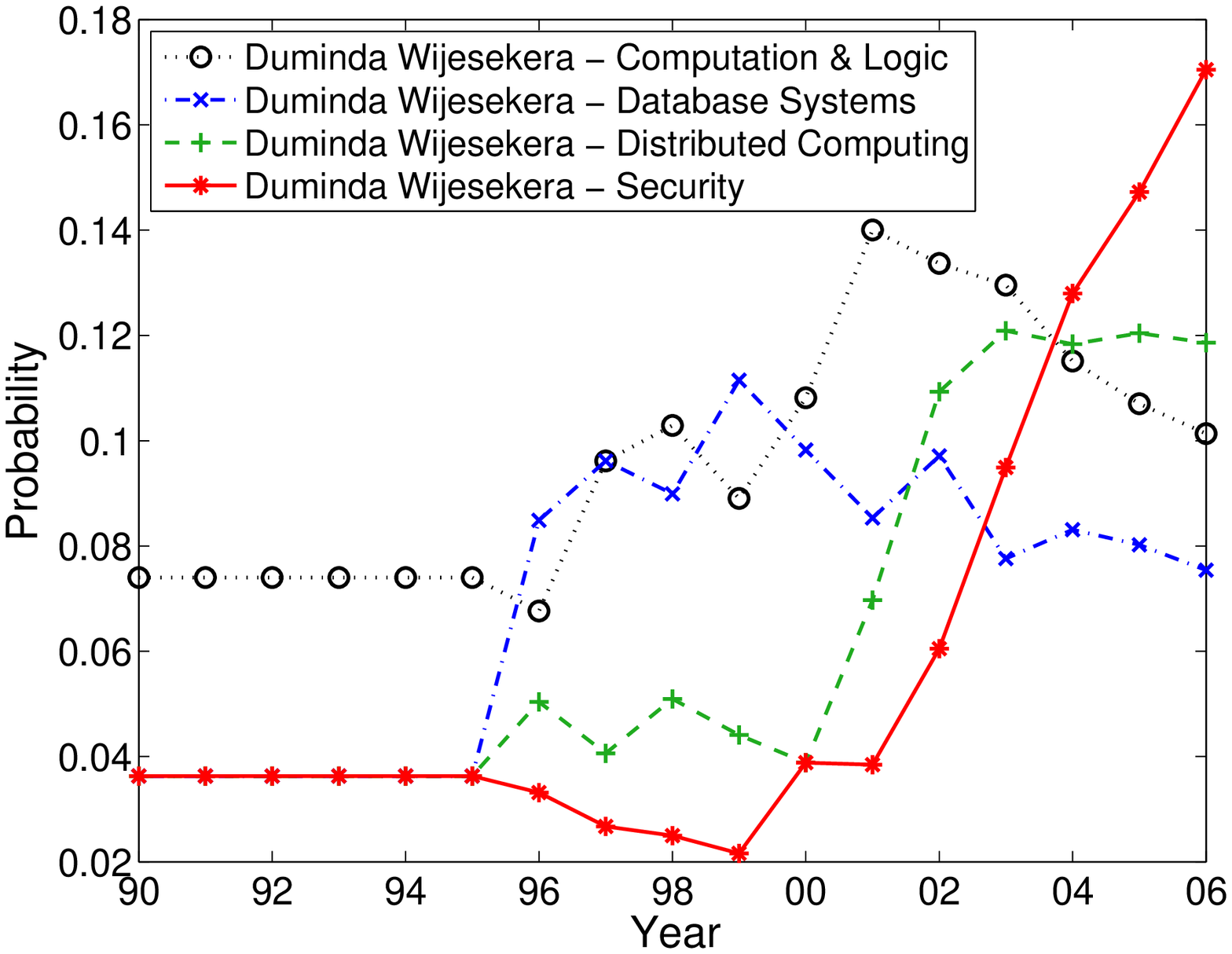}
    \label{fig:duminda_wijesekera_lda_topics}
  }
  \subfigure[Co-Authors' Computation \& Logic Topic]
  {
    \includegraphics[width=2.2in]{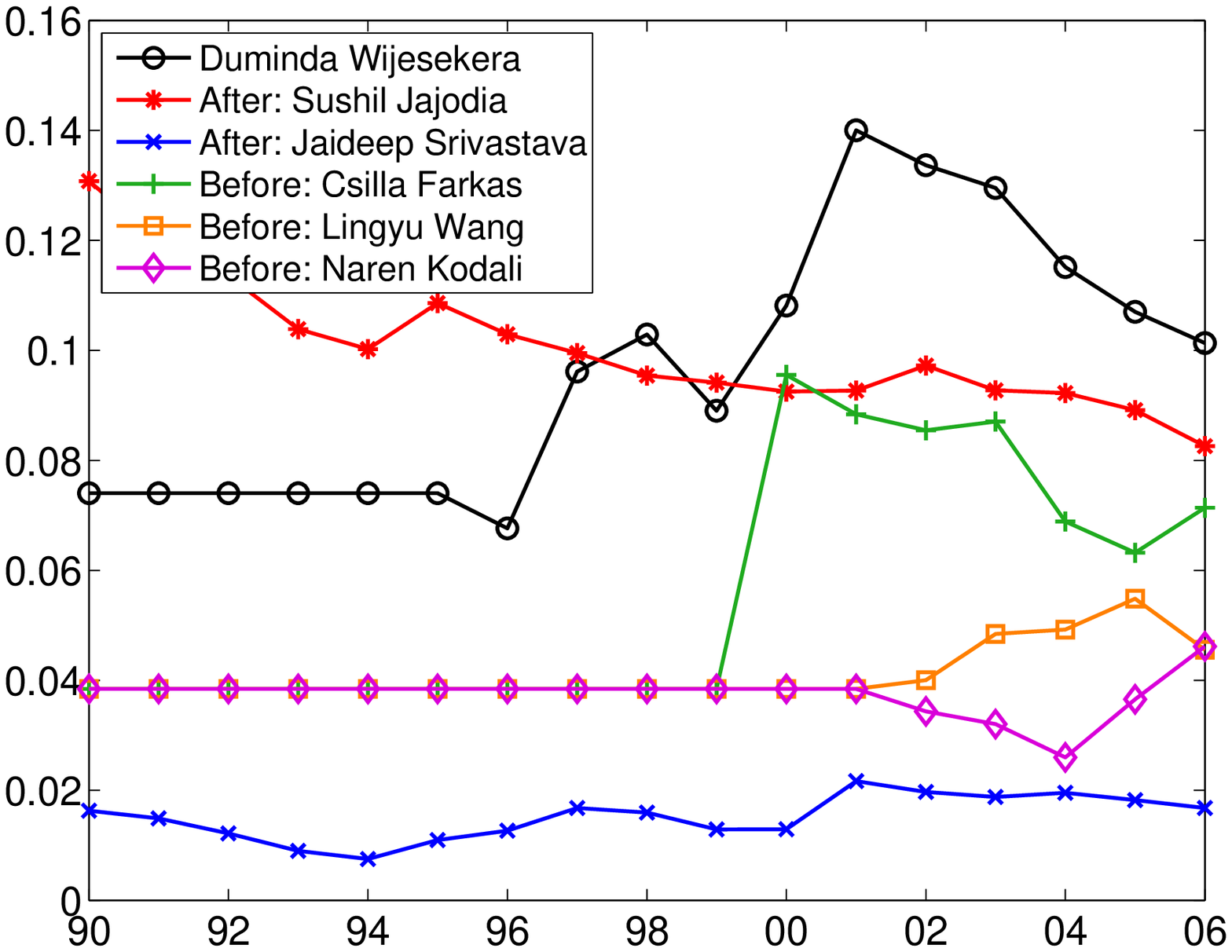}
    \label{fig:duminda_wijesekera_coauthors_lda_topics}
  }
  \caption{LDA Results: Duminda Wijesekera's and His Co-Authors' Topic Interests from 1990 to 2006}
  \label{fig:duminda_and_coauthor_lda_topics}
\end{figure}

Figure \ref{fig:duminda_wijesekera_lda_topics} shows the LDA-induced topic distribution of D. Wijesekera from year 1990 to 2006 (also the top four topics), and Table \ref{chap:gc:tbl:lda_topics} shows the corresponding topic words. Instead of the ``Security'' topic, the ``Computation \& Logic'' topic occupies the largest area here.

We then analyzed the topic proportions of the ``Computation \& Logic'' topic for D. Wijesekera's co-authors, as given in Figure \ref{fig:duminda_wijesekera_coauthors_lda_topics}. Unlike the previous LDTM case, we could not find any co-authors who have significant correlation with D. Wijesekera in the ``Computation \& Logic'' topic. This indicates that the accuracy of the topic distributions are important for us to infer the Granger causality between the authors. 

The LDTM model is able to appropriately decay the importance of other topics and focus on emergence of new topics such as ``Security'' in D. Wijesekera's case. Granger causality would then allow us to find co-authors who are socially correlated to D. Wijesekera in order to explain the emergence or change of academic interests over time.

\begin{table}[!htb]
  \caption{Topics Derived from LDA}
    \label{chap:gc:tbl:lda_topics}
    \begin{tabular} {|c|c|c|c|}
      \hline
      \textbf{Security} & \textbf{Database Systems} & \textbf{Logic \& Computation} & \textbf{Distributed Computing} \\
      \hline
      security & video & logic & service \\
      secure & peer & programming & management \\
      scheme & multimedia & verification & grid \\
      efficient & adaptive & languages & computing \\
      privacy & content & formal & mobile \\
      \hline
    \end{tabular}
\end{table}

\subsection{Knowledge Discovery using Temporal Social Correlation}

We now show the application of topic distributions as time series for computing the $TSC$ between two authors $i$ and $j$ at a time point $\tau$. The parameters ``width'' and ``lookahead'' are both set as 4. Using the co-authorship information in our sampled \emph{ego-3} datasets, we choose some pairs of authors and the year of publication at time point $\tau$ in order to compute $TSC$. 
We formulate the following three hypotheses:
\begin{enumerate}
  \item \textbf{AB}: If $j$ is the first author and $i$ is the second author of a publication written at $\tau$, then $i$ transfers information to $j$, i.e. $TSC(i \rightarrow j, \tau) > TSC(j \rightarrow i, \tau)$.
  \item \textbf{AZ}: If $j$ is the first author and $i$ is the last author of a publication written at $\tau$, then $i$ transfers information to $j$, i.e. $TSC(i \rightarrow j, \tau) > TSC(j \rightarrow i, \tau)$.
  \item \textbf{Bf\_Af}: If $j$ and $i$ are authors of a publication written at $\tau$ with more than two authors and $j$ comes before $i$, then $i$ transfers information to $j$, i.e. $TSC(i \rightarrow j, \tau) > TSC(j \rightarrow i, \tau)$.
\end{enumerate}
and compute a $Ratio$ metric for each scenario, defined as:
\[ Ratio = \frac{\sum_{(i,j) \in P} I(TSC(i \rightarrow j, \tau) > TSC(j \rightarrow i, \tau))}{|P|} \]
where $I(.)$ is the indicator function, and $P$ is the set of all user pairs $(i, j)$ considered in the evaluation. In other words, the $Ratio$ may be interpreted as the proportion (or probability) of the $TSC$ from user $i$ to $j$ being greater than that from $j$ to $i$.

To prevent confounding factors in our experiments, \emph{we excluded the publications that have the authors' last name arranged in ascending order}. We removed 41\% from ACMDL and 44\% from DBLP, although some papers might have authors last name in ascending order due to coincidence.

\begin{figure}[!htb]
  \centering
  \subfigure[ACMDL: $Ratio$]
  {
    \includegraphics[width=2.2in]{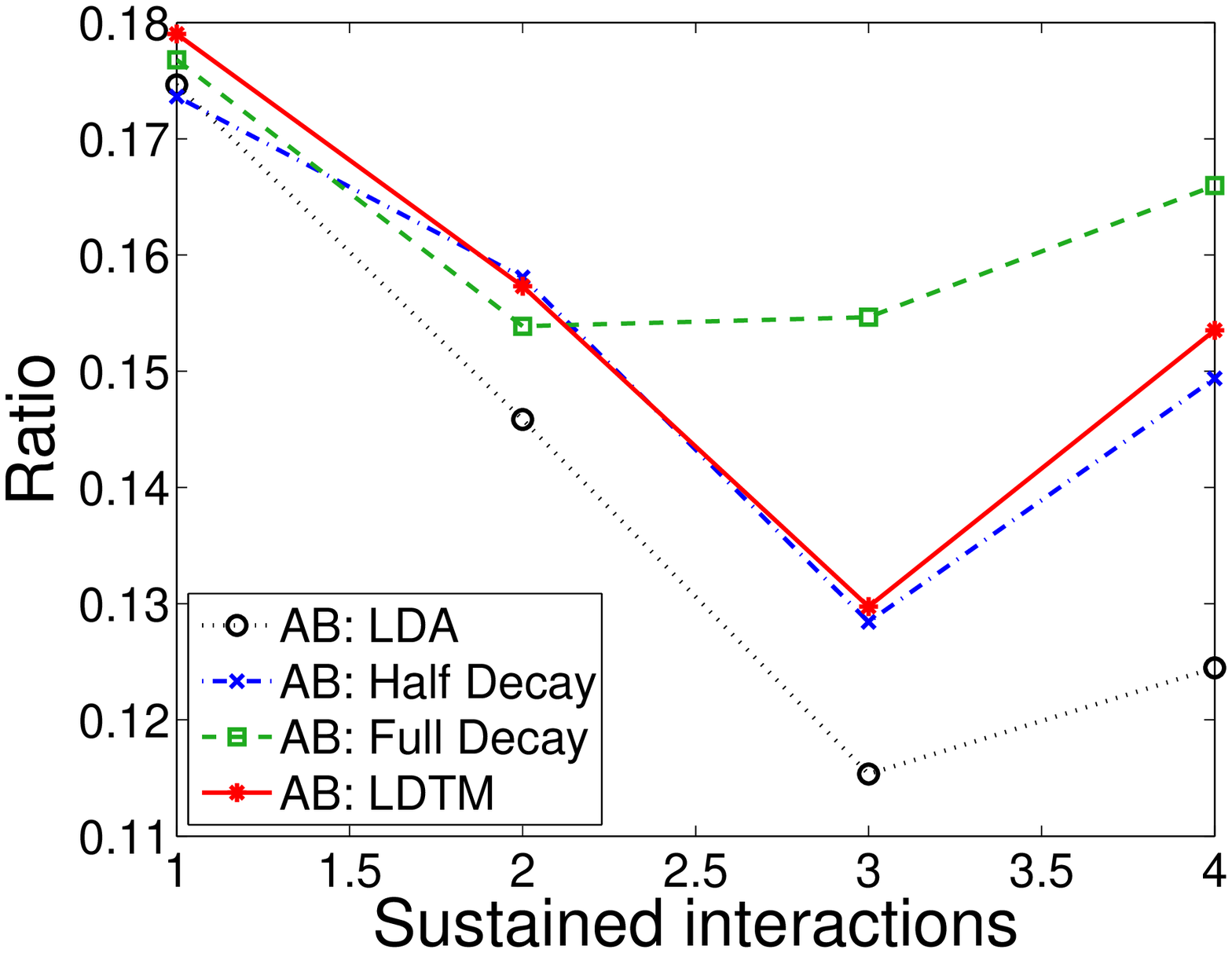}
    \label{fig:acmdl_ab_ba_gc_timeline}
  }
  \subfigure[DBLP: $Ratio$]
  {
    \includegraphics[width=2.2in]{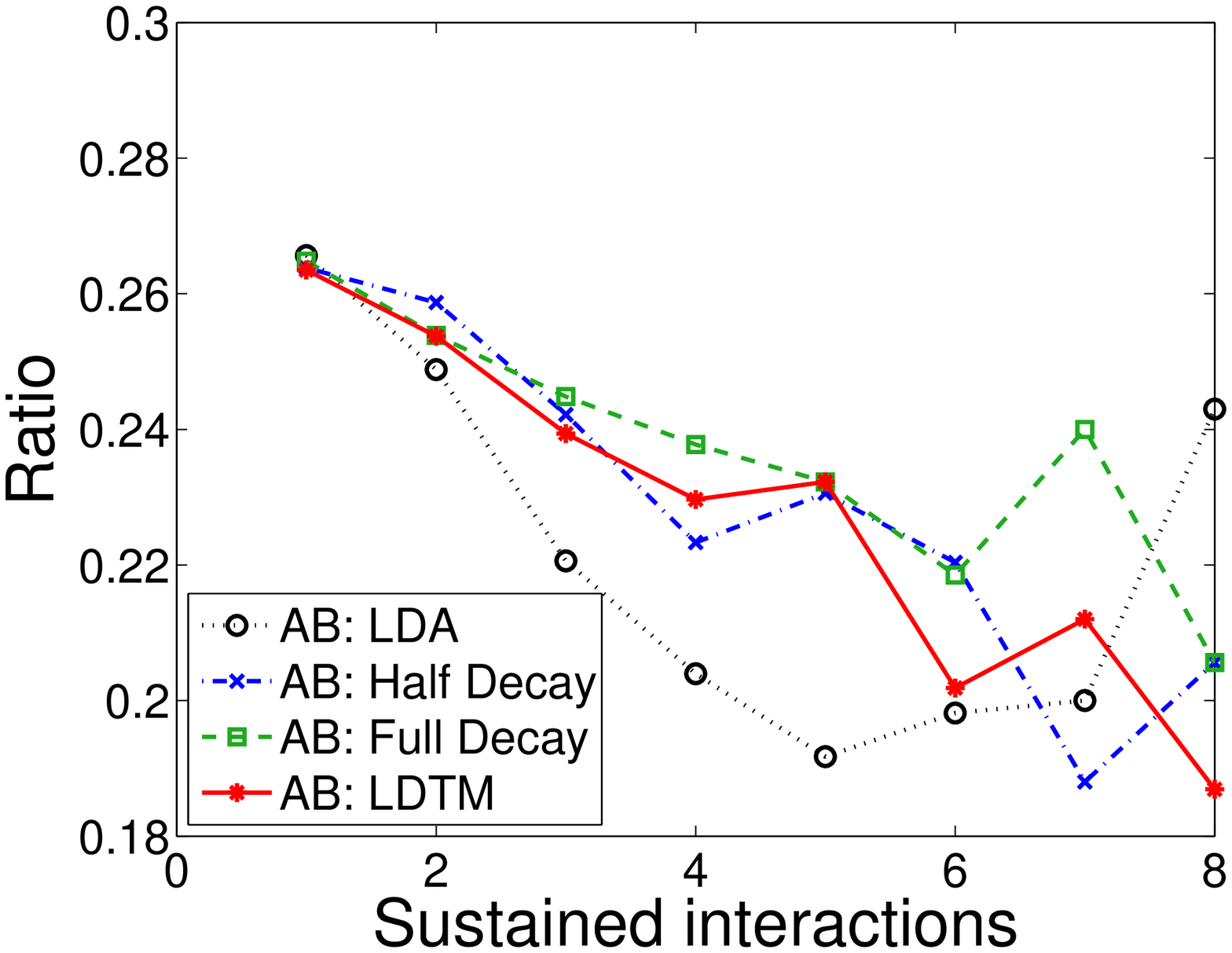}
    \label{fig:dblp_ab_ba_gc_timeline}
  }
  \caption{AB: Ratio based on Time Series of Various Models}
  \label{fig:ab_ba_gc}
\end{figure}

\begin{figure}[!htb]
  \centering
  \subfigure[ACMDL: $Ratio$]
  {
    \includegraphics[width=2.2in]{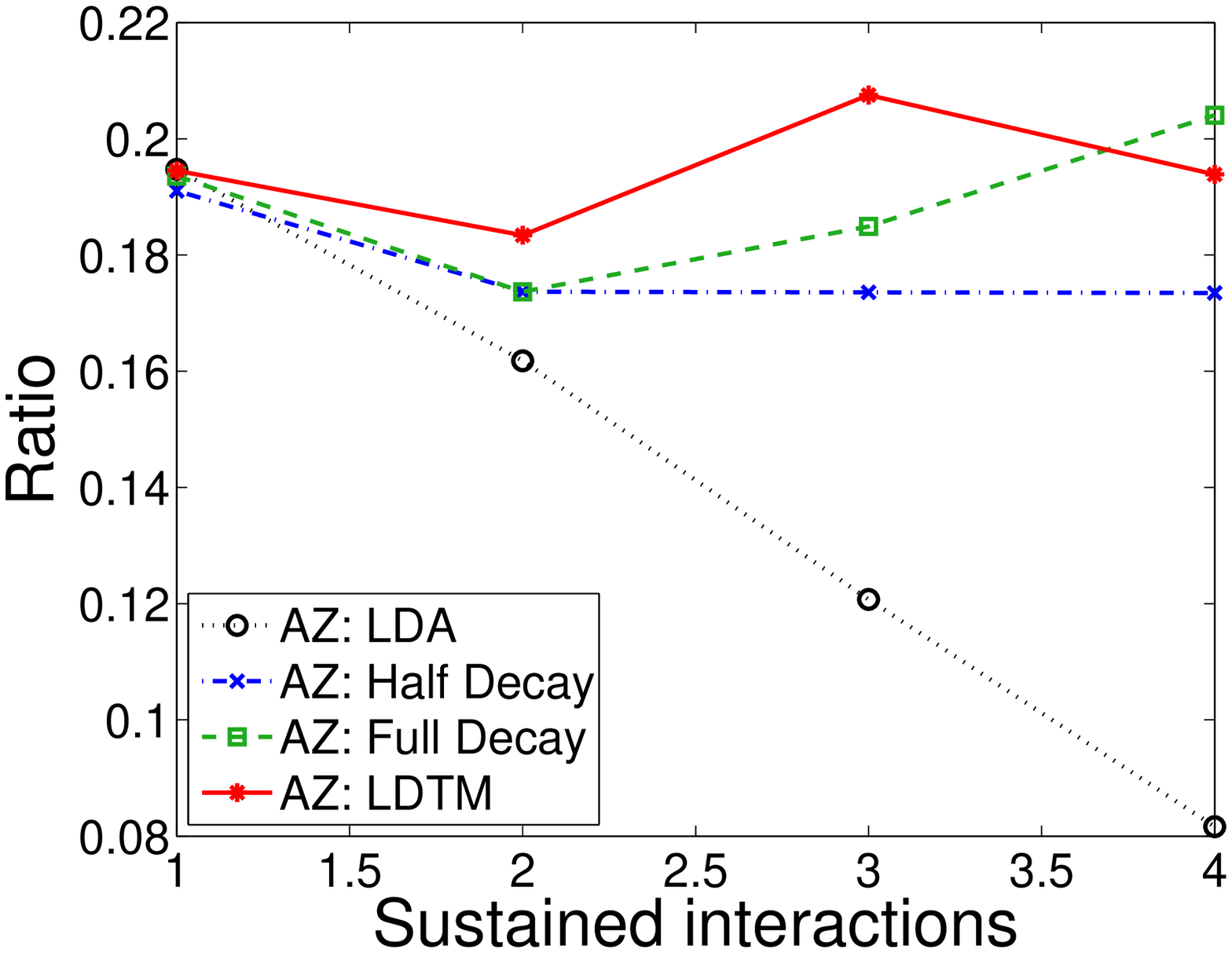}
    \label{fig:acmdl_az_za_gc_timeline}
  }
  \subfigure[DBLP: $Ratio$]
  {
    \includegraphics[width=2.2in]{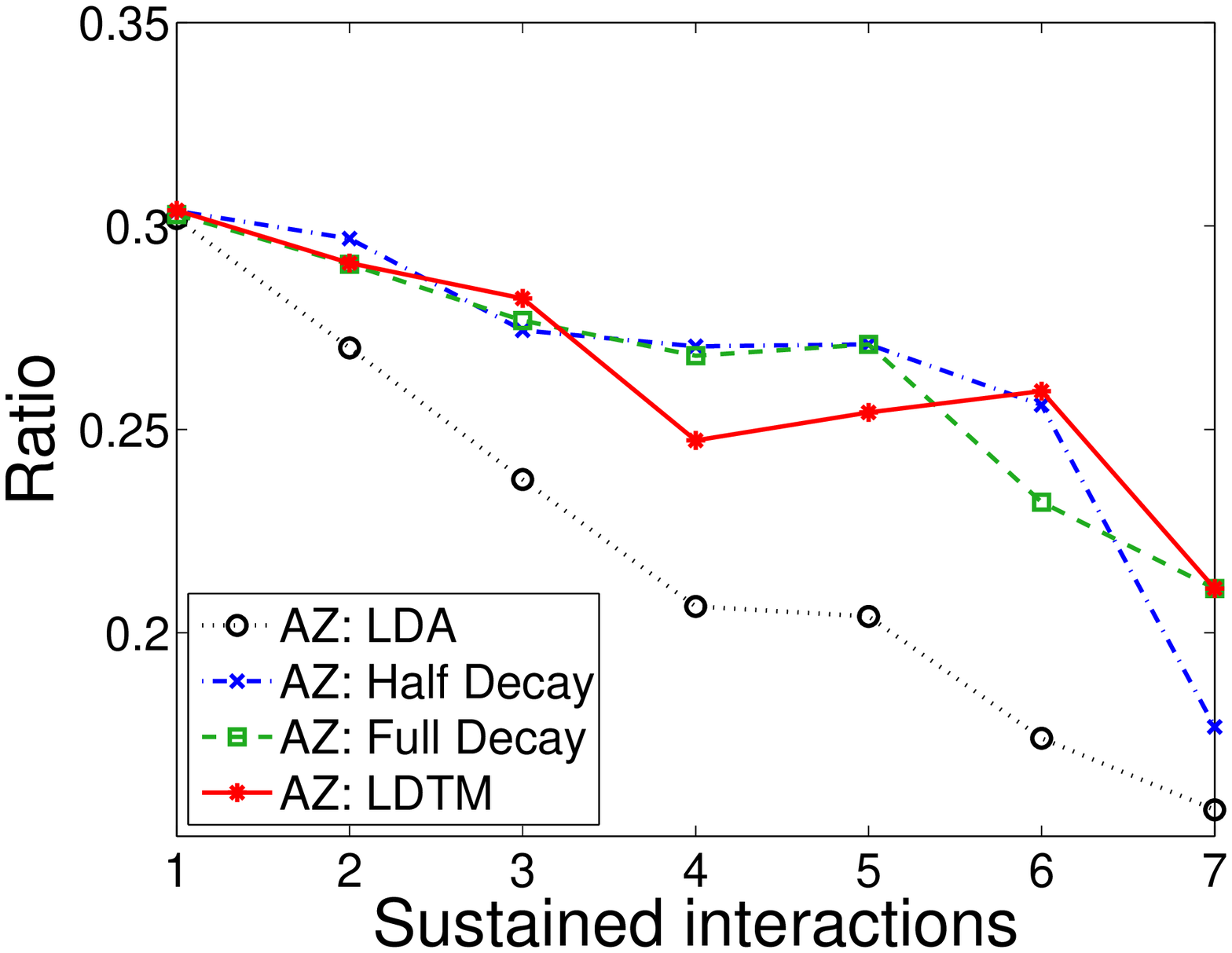}
    \label{fig:dblp_az_za_gc_timeline}
  }
  \caption{AZ: Ratio based on Time Series of Various Models}
  \label{fig:az_za_gc}
\end{figure}

\begin{figure}[!htb]
  \centering
  \subfigure[ACMDL: $Ratio$]
  {
    \includegraphics[width=2.2in]{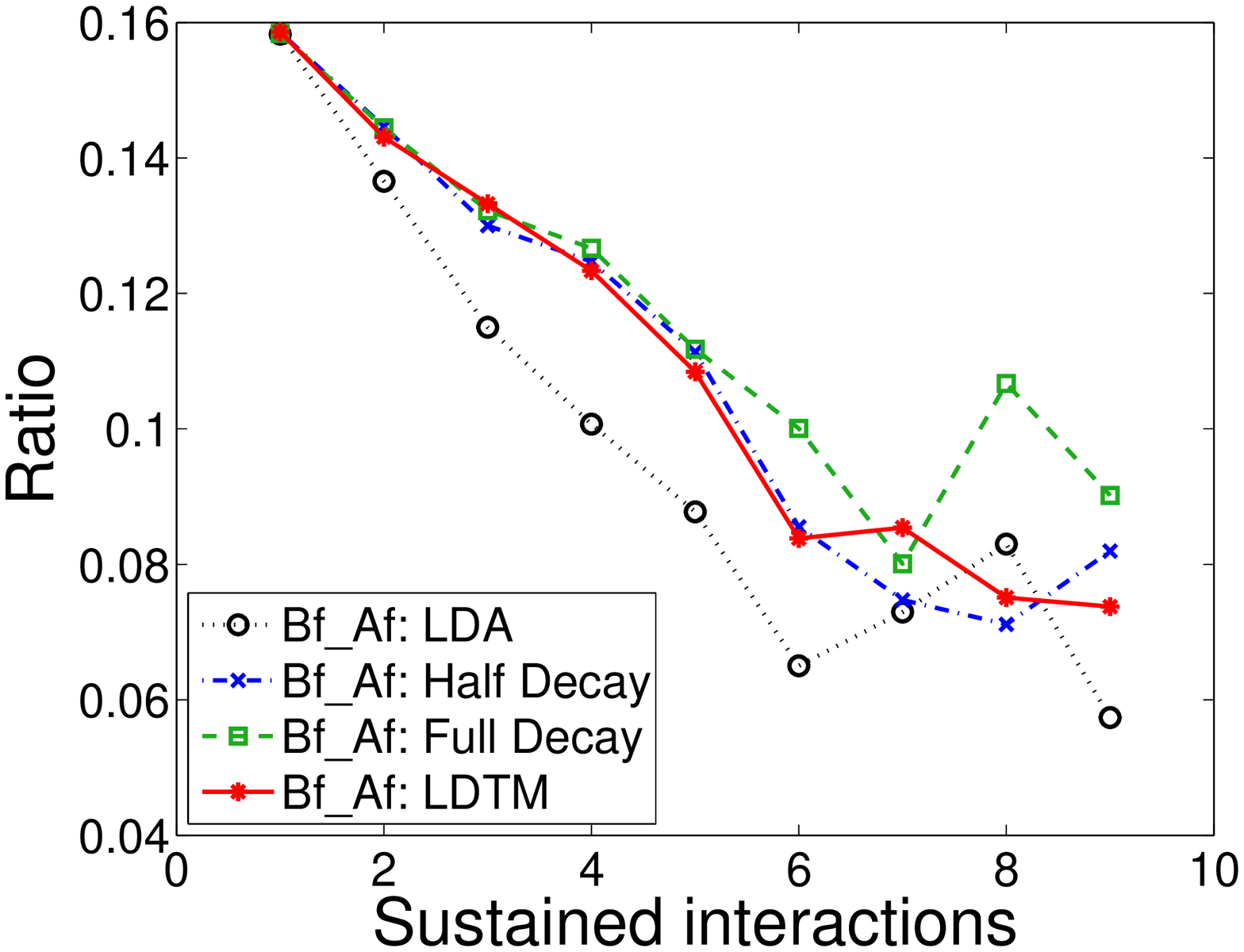}
    \label{fig:acmdl_bef_aft_gc_timeline}
  }
  \subfigure[DBLP: $Ratio$]
  {
    \includegraphics[width=2.2in]{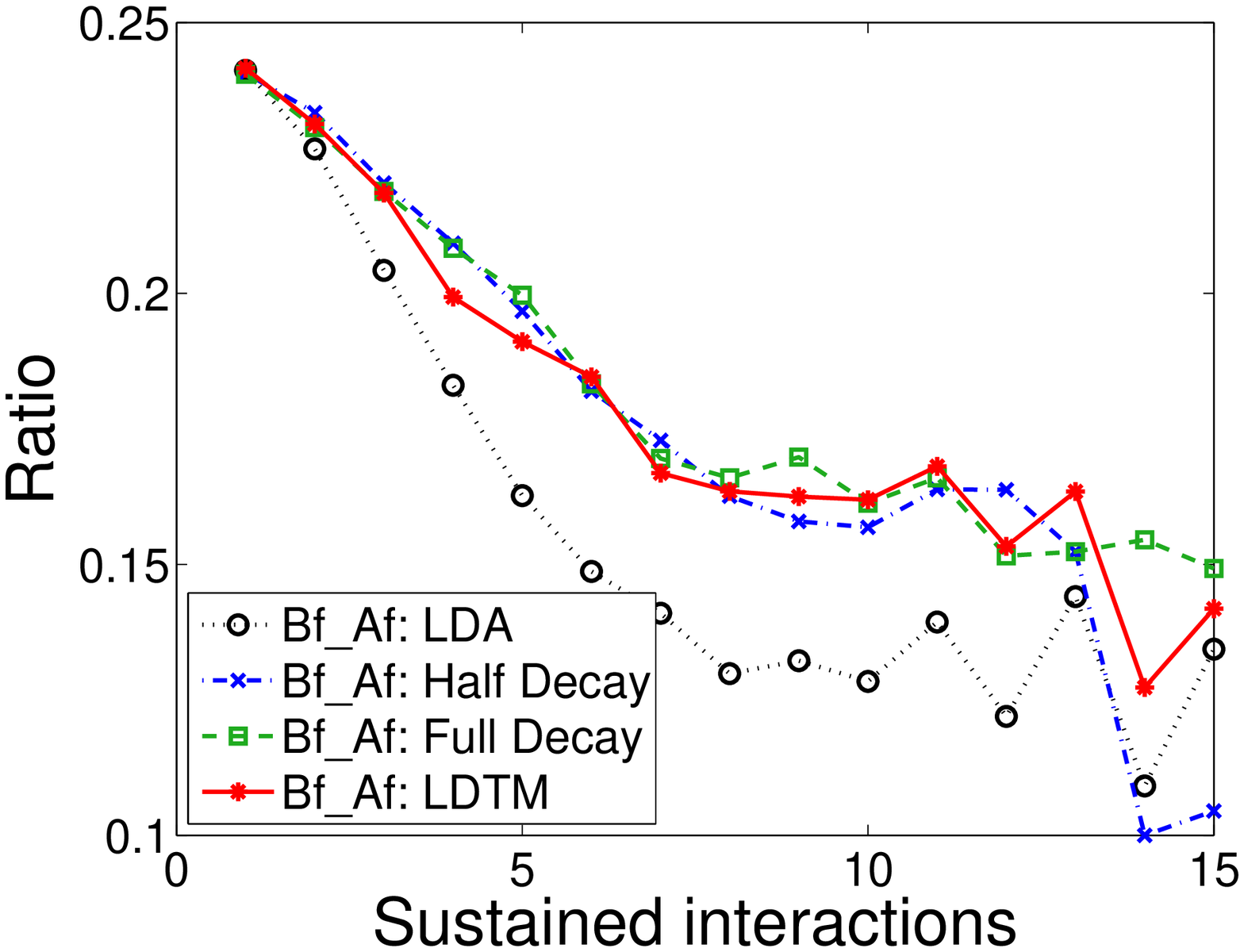}
    \label{fig:dblp_bef_aft_gc_timeline}
  }
  \caption{Bf\_Af: Ratio based on Time Series of Various Models}
  \label{fig:bef_aft_gc}
\end{figure}

For every $i, j$ author pair who co-authored at least once, we computed the $Ratio$ for each time step $\tau$. We then analyzed the $Ratio$ values for every pair with respect to the number of time steps in which they had sustained the co-author relationship. Figures \ref{fig:ab_ba_gc} to \ref{fig:bef_aft_gc} show the $Ratio$ values versus the number of time steps the author pairs have sustained their co-authorships. The subplots (a) of Figures \ref{fig:ab_ba_gc} to \ref{fig:bef_aft_gc} present the $Ratio$ values for ACMDL, while subplots (b) of Figures \ref{fig:ab_ba_gc} to \ref{fig:bef_aft_gc} give the $Ratio$ values for DBLP. Only bins with more than 90 data points are shown in the Figures.

Figures \ref{fig:ab_ba_gc} to \ref{fig:bef_aft_gc} reveal several interesting phenomena. To describe the phenomena, we use the notation $I$ to denote the set of authors whose names are placed at the back, and $J$ to denote the set of authors whose names are placed in front. In general, since we have earlier filtered out the papers with alphabetical ordering, researchers ($J$) who do the bulk of the research have their names placed in front of their co-authors ($I$).
\begin{enumerate}
  \item All models and hypotheses show that the ratio is always below 0.5. This indicates $I$ does not necessarily influence $J$. On the contrary, the results show a high probability of $J$ influencing $I$, i.e. Researchers who do the most work influence their co-authors.
  \item The ratio is always on a downtrend which indicates that the influence $I$ has on $J$ decreases over time. As researchers progress in their research career, the influence their co-authors have on them decreases over time.
  \item By comparing between the Ratio of Figures \ref{fig:ab_ba_gc} and \ref{fig:bef_aft_gc}, the second author $i$ has more influence on the first author $j$, as compared to the generic case where $j$ can be any author (e.g. 2nd, 3rd, etc.) and $i$ comes after $j$.
  \item By comparing between the Ratio of Figures \ref{fig:ab_ba_gc} and \ref{fig:az_za_gc}, the last author $i$ influence the first author $j$, more than the second author.
\end{enumerate}

Based on these results, we can conclude that, for pairs of authors who wrote a paper together, it is highly likely that the author whose name appears in front (Granger) causes those whose names appear at the back to change their topic distributions over time. We therefore reject the earlier hypotheses of AB, AZ and Bf\_Af.

\section{Conclusion}
\label{chap:gc:sec:summary}

This paper presents a means for identifying Temporal Social Correlation (TSC) based on latent topic representation of item adoptions that users perform over time. We propose a Linear Dynamical Topic Model (LDTM) that synergizes the merits of probabilistic topic models and Linear Dynamical Systems (LDS) in order to capture users' adoption behavior over time. The EM algorithm for solving the model draws upon Gibbs Sampling and Kalman Filter for inference in the E-Step, followed by the M-Step which minimizes the Kullback-Leibler (KL) divergence between the prior and posterior distributions for estimating the dynamics matrix. By taking into account both the stability and non-negativity constraints, we derive a dynamics matrix that represents how users decay their adoption behaviors and preferences over time.

By using the users' topic distributions at different time steps, we construct each user's time series and compare it with their co-authors' using Granger-causal tests. Our experiments on bibliographic datasets demonstrate that, by employing Granger causality on the time series, we can calculate the TSC between authors of the paper and discover that the ordering of authors' names plays a role in how information transfer among them.


Ultimately, all the measurements we made is to further the science of predicting the future. However, it remains to be seen whether quantifying social influence through TSC could be used for making recommendations to users on what items to adopt, and that could be a future direction in our work.






\bibliographystyle{abbrv}
\bibliography{references}



\end{document}